\documentclass{elsart}
\usepackage{graphicx}
\usepackage{amsmath}
\usepackage{amssymb}
\def\Vec#1{\mbox{\boldmath $#1$}}
\input epsf.sty
\def\PLA{{Phys. Lett.} A}

\def\PRL{Phys. Rev. Lett.}
\def\IEEE{IEEE Transactions on Circuits and Systems, CAS-}
\usepackage{amssymb}

\begin{document}
\begin{frontmatter}
\title
{Shil'nikov Chaos control using Homoclinic orbits and the Newhouse region}
\author
{Sadataka Furui and Shohei Niiya} 
\address
{Graduate School, Teikyo University, \\1-1 Toyosatodai, Utsunomiya 320-8551, Japan}
\ead{furui@umb.teikyo-u.ac.jp}

\begin{abstract}
A method of controlling Shil'nikov's type chaos using windows that appear in the
1 dimensional bifurcation diagram when perturbations are applied, and using existence of stable homoclinic orbits near the unstable one is presented and applied to the electronic Chua's circuit. A demonstration of the chaos  control in the electronic circuit experiments and their simulations and bifurcation analyses are given. 
\end{abstract}

\begin{keyword}
Shil'nikov Chaos \sep homoclinic orbit \sep Chua's circuit
\PACS 5.45 \sep 47.52 \sep 47.27 
\end{keyword}
\end{frontmatter}

\section{Introduction}

Controlling chaos is important for application in the comunication system\cite{hgo93,hgom94} and controlling dynamics in technology like convective flows, laser excitation and electronic currents. There are two well-known methods for controlling chaos 1) OGY method\cite{ogy90} and  2) feedback (Pyragas) method\cite{pyr92}.  In the case of electronic current, chaotic oscillation of Shil'nikov type\cite{shil65} occurs in the Chua's circuit and the chaotic patterns depending on the parameters  of the dynamical system were extensively investigated\cite{CKM86}. 

Chua's circuit consists of a autonomous circuit which contains three-segment piecewise-linear resistor, two capacitors, one inductor and a variable resistor.
The equation of the circuit is described by
\begin{equation}\label{chua_bare}
\left\{\begin{array}{l}
C_1\frac{d}{dt}v_{C1}=G(v_{C2}-v_{C1})-\tilde g(v_{C1},m_0,m_1)\\
C_2\frac{d}{dt}v_{C2}=G(v_{C1}-v_{C2})+i_L\\
L\frac{d}{dt}i_L=-v_{C2}
\end{array}\right.
\end{equation}
where $v_{C1}$ and $v_{C2}$ are the voltages of the two capacitors (in V), $i_L$ is the current that flows in the inductor (in A), $C_1$ and $C_2$ are capacitance (in F), $L$ is inductance (in H) and $G$ is the conductance of the variable resistor (in $\Omega^{-1}$). The three-segment piecewise-linear resistor which constitutes the non-linear element is characterized by
\[
\tilde g(v_{C1},m_0,m_1)=(m_1-m_0)(|v_{C1}+B_p|-|v_{C1}-B_p|)/2+m_0 v_{C1}, 
\]
where $B_p$ is chosen to be 1V, $m_0$ is the slope (mA/V) outside $|v_{C1}|>B_p$  and $m_1$ is the slope inside $-B_p<v_{C1}<B_p$.  
The function $\tilde g(v_{C1},m_0,m_1)$ can be regarded as an active resistor. If it is locally passive the circuit is tame but when it is locally active it keeps supplying power to the external circuit\cite{MCK85}. The chaotic behavior is expected to be due to the power dissipated in the passive element. A difference from the van-der Pol circuit is that the passive area and the active area are not separated by the strange attractor, but they are entangled.

 We control chaotic circuits by triggering simple electronic pulse as \cite{SN03}. Main difference from this work is that we choose a proper height of the pulse, so that the periodic trajectory is kept and we do not need to reset the current. 

By coupling two Chua's circuit, one can control the chaotic motion by phase synchronization\cite{drmc03}. We adopt the phase synchronization technique  
 on a single Chua's oscillator and control the chaotic motion.
We adopt also the feedback control technique to the Chua's oscillator.

\begin{figure}[htb]
\begin{center}
\epsfysize=220pt\epsfbox{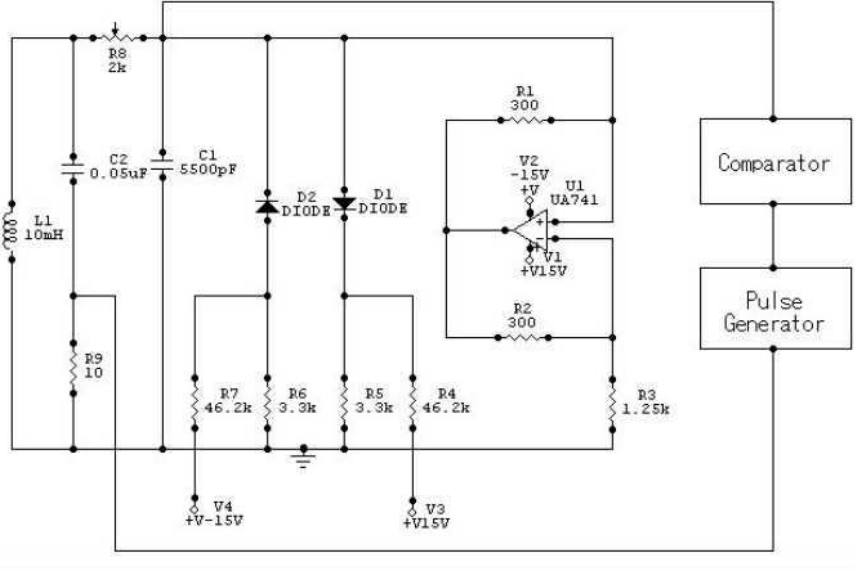}
\end{center}
\caption{Chua's electronic circuit with impulse generator.}\label{chua_circuit}
\end{figure}

The Chua's electronic circuit and connection of the comparator and the pulse generator are shown in Fig.\ref{chua_circuit}.
Changing the scales of variables as $C_i\times 10^7, L\times 20, G\times 10^3, m_i\times 10^3$ and $t\times 2\times 10^4$, we obtain the differential equation 
$\displaystyle \frac{d}{dt}{\bf x}={\bf f}\cdot{\bf x}$ where ${\bf x}=(x(\tau),y(\tau),z(\tau))$ as
\begin{equation}\label{chua_eq}
\left\{\begin{array}{l}
\frac{d}{d\tau}x=\alpha(y-x-g(x,m_0,m_1))\\
\frac{d}{d\tau}y=x-y+z\\
\frac{d}{d\tau}z=-\beta y
\end{array}\right.
\end{equation}
Here $\displaystyle \alpha=\frac{C_2}{C_1}, \beta=\frac{C_2}{L G^2}$ and 
\[
g(x,m_0,m_1)=(m_1-m_0)(|x+1|-|x-1|)/2+m_0 x.
\]
We choose $C_1 = {1\over 9}, C_2 = 1, L = {1\over 7}, G = 0.7$, $m_0 = -0.5,  m_1 = -0.8, B_p = 1$. 

The differential equation is invariant under symmetry $(x,y,z)\to(-x,-y,-z)$ and the state space can be divided into three domains by the planes $U_{-1}=\{z=-1\}$ and $U_+=\{z=1\}$. In each domain there are equilibrium points at
\begin{tabular}{ccc}
$P_+=(-k,0,k)$& O=(0,0,0) & $P_-(k,0,-k)$\\
\end{tabular}
where $\displaystyle k=\frac{m_0-m_1}{m_0+1}$. In the three regions, the linearized ${\bf f}$ has a real eigenvalue $\gamma$ and a pair of complex eigenvales $\sigma\pm i\omega$.  

Shil'nikov considered a continuous vector field $X$ on ${\bf R}^3$ with equilibrium points $p$ such that $|\sigma|<|\gamma|$ and  $\omega\ne 0$ and that there is a homoclinic orbit from $p$. Then he proved that there is a perturbation $Y$ of $X$ such that $Y$ has invariant sets containing transversal homoclinic orbits\cite{GH83,shil65}.
It means if the Chua's circuit satisfies the condition $|\sigma|<|\gamma|$ the trajectory can be perturbed into an orbit with countable sets of horseshoe, or  Chaos is embedded in the system.  

Newhouse showed, in addition to bifurcations mechanism of Shil'nikov, subtle and complicated dynamical behaviour associated with homoclinic tangency exists in the diffeomorphism of continuous vector field $X$ on ${\bf R}^2$. He defined a zero-dimensional hyperbolic invariant set  $\Lambda$ in a plane, stable manifold $W^s(\Lambda)$, unstable manifolds $W^u(\Lambda)$, thickness of $W^s(\Lambda)$ as $\tau^s(\Lambda)$ and that of $W^u(\Lambda)$ as $\tau^u(\Lambda)$. He showed when 1) $\tau^s(\Lambda_2)\cdot \tau^u(\Lambda_1)>1$, 2) 
$W^s(\Lambda_2)\cap W^u(\Lambda_1)$ and $W^u(\Lambda_2)\cap W^s(\Lambda_1)$ both have points of transversal intersection which are not in $\Lambda_1\cup\Lambda_2$, 3)$W^s(\Lambda_2)$ and $W^u(\Lambda_1)$ have a point of tangency, then the diffeomorphism has infinite number of sinks\cite{new79,GH83}.

The Shil'nikov's chaos occurs in the ${\bf R}^3$ but the theorem on 2-dimensional(2-d) Cantor set would apply also to 3-d Cantor set since the homoclinicity of the orbit is the same.

A sufficient condition for finding a map $\bf f$ in the neighborhood of $p$ such that for $m>M$, ${\bf f}^m$ has an invariant set $\Lambda_m$ topologically equivalent to the horseshoe is that the eigenvalues of the Jacobi matrix at the fixed point $\rho$ and $\lambda$ satisfy $\rho<1<\lambda<\rho^{-1}$\cite{GH83}.  When $\tau^s(\Lambda_2)\cdot \tau^u(\Lambda_1)>1$ is satisfied and $W^s(\Lambda)$ has a point of tangential intersection with $W^u(\Lambda)$, $\Lambda$ is called wild hyperbolic set and the region around this set is called Newhouse region.

In the windows region of the 1-d bifurcation diagram as a function of the strength of the perturbation, there are countable periodic orbits belonging to $W^s(\Lambda_2)$ and to $W^u(\Lambda_1)$. Since it is plausible that there are points of transversal intersection outside $\Lambda_1\cup\Lambda_2$ in the windows region, perturbation in this region would kick the unstable orbit to one of the sinks.
We define the strength of the pulse or the amplitude of synchronizing oscillation or the amplitude of the feedback by using the information of the position of windows in the 1-d mapping of the bifurcation diagram. Various kinds of shift from unstable manifold to stable manifold occurs due to the dense presence of stable homoclinic orbits near each unstable orbits in the Newhouse region\cite{GH83}.

This paper is organized as follows. In sect.2 we explain the experimental setup of the Chua's circuit control system and chaotic patterns. 
 The method of controlling the double scroll and the experimental results are shown in sect.3. In sect.4 we show the method of bifurcation analysis. Impulse control of the double scroll is analyzed in sect.5. Drive control and feedback control are analyzed in sect.6 and in sect.7, respectively. Analysis of the spiral control is given in sect.8.  Discussion and outlook are presented in sect.9.

\section{The Chua's circuit}

\begin{figure}[htb]
\begin{minipage}[b]{0.47\linewidth}
\begin{center}
\epsfysize=120pt\epsfbox{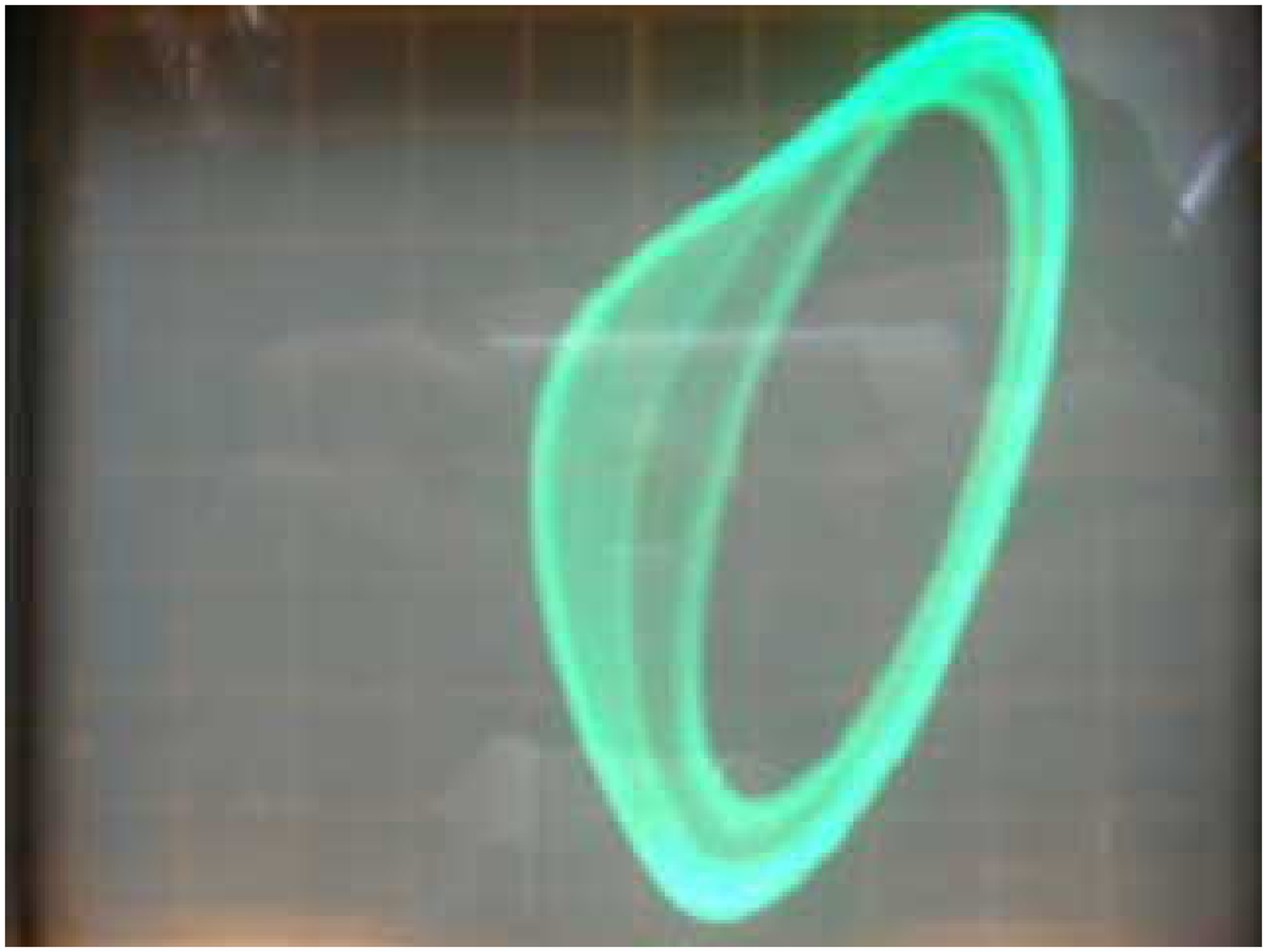}
\end{center}
\caption{The Chua's spiral.}\label{chua_spiral}
\end{minipage}
\hfill
\begin{minipage}[b]{0.47\linewidth}
\begin{center}
\epsfysize=120pt\epsfbox{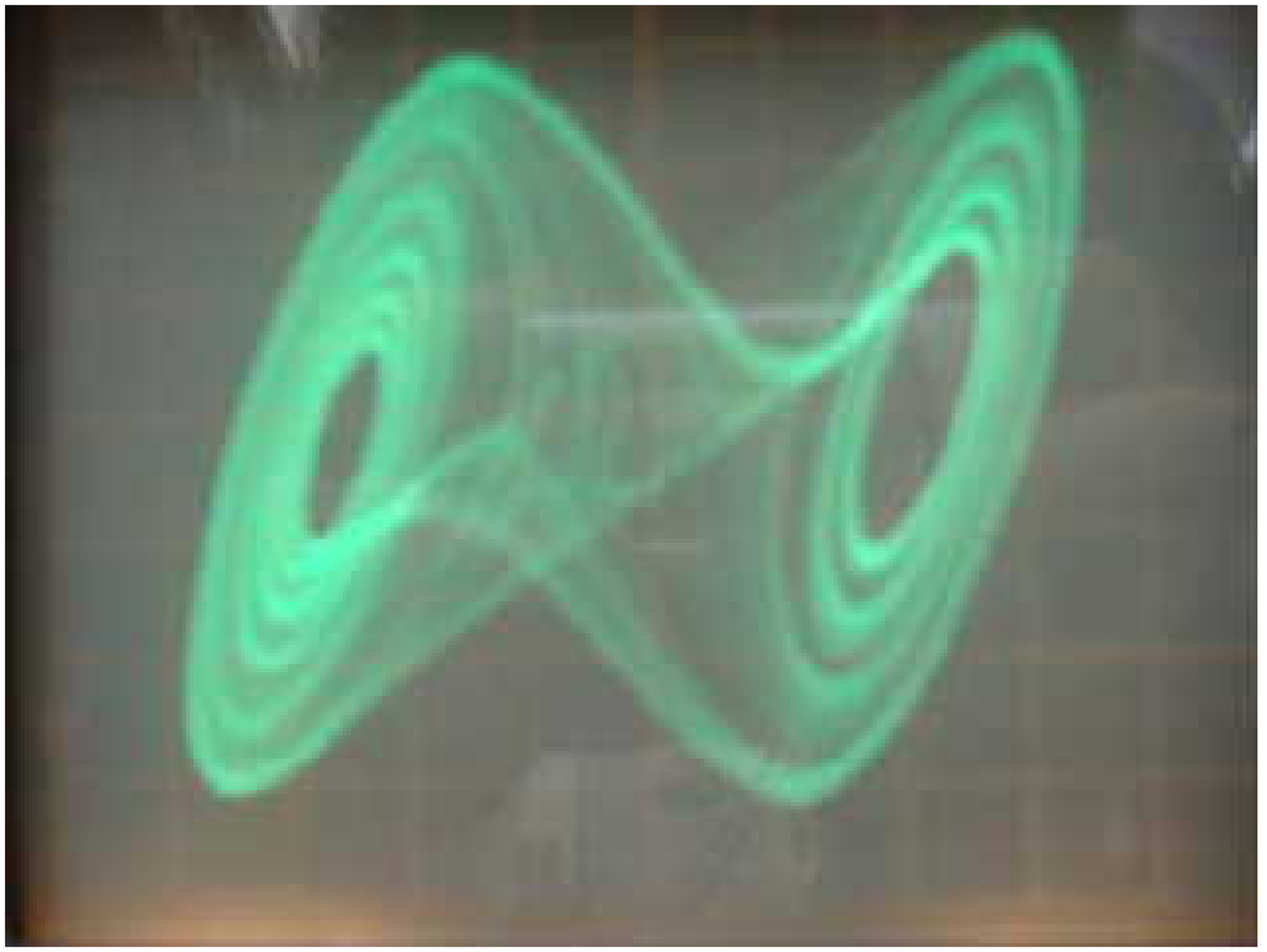}
\end{center}
\caption{The Chua's double scroll.}\label{chua1_2}
\end{minipage}
\end{figure}

The dynamical system of the Chua's circuit is defined as eq.(\ref{chua_eq}).
When the variable resistance is about $R=1616\Omega$, on the oscilloscope displaying $v_{C1}-v_{C2}$, a circle appears and as the  resistence is reduced to $R=1534\Omega$ suddenly a chaotic spiral as shown in Fig.\ref{chua_spiral} appears. When $R$ is reduced to $R=1528\Omega$ a 3 cycle spiral  appears and when $R$ is 1457$\Omega$ a double scroll as shown in Fig.\ref{chua1_2} appears.
When $R$ is further reduced to 1402$\Omega$, a periodic double scroll with 3 cycles around each fixed point appears and for smaller $R$ the double scroll disappears. By measuring Lyapunov exponents we check that the spiral and the double scroll are chaotic, and the chaotic behavior is implied by the presence of the period-3 cycle according to the Li-Yorke's theorem\cite{GH83}. 

Eigenvalues of the Jacobi matrix $\displaystyle \frac{\partial {\bf f}}{\partial {\bf x}}$ where ${\bf x}=(x,y,z)$ for $|x|<1$ are defined as $\sigma_0\pm i\omega_0, \gamma_0$ and for $x<-1, 1<x$ are defined as $\sigma_1\pm i\omega_1, \gamma_1$. Numerical values as a function of the resistence $R$ used in the experiment and the conductance $G$ used in the simulation are shown in Table 1. The value $G$ used in the simulation does not necessarily coincide with the experimental value $1000/R$. The Shil'nikov's condition for the homoclinic chaos is $|\sigma|<|\gamma|$ and $\omega\ne 0$ and the condition is satisfied in the range of $0.55<G<0.643$.

\begin{table}\label{eigenvalue}
\begin{center}
\caption{Eigenvalues of the Jacobi matrix at the fixed point.}
\begin{tabular}{cc|ccc|ccc|c}
$R(\Omega)$ & $G$& $\sigma_0$ & $\omega_0$ & $\gamma_0$ &$\sigma_1$ & $\omega_1$ & $\gamma_1$& \\ 
\hline
1635 & 0.55& -0.99 & 3.52 & 5.11 & -0.06 & 2.84 & -1.70 &\\
1616 & 0.562 &-1.00 & 3.40 & 4.85 & 0.02 & 2.79 & -2.03 & 1 cycle\\
1549 & 0.5645 &-1.00 & 3.38 & 4.80 & 0.03 & 2.79 & -2.10 & 2 cycle\\      
1540 & 0.5653&-1.00 & 3.37 & 4.77 & 0.03 & 2.78 & -2.12 & 4 cycle\\
1534 & 0.566&-1.00 & 3.36 & 4.76 & 0.04 & 2.78 & -2.14 & spiral\\
1528 & 0.56638&-1.00 & 3.36 & 4.76 & 0.04 & 2.78 & -2.15 & 3 cycle\\
1457 & 0.585&-1.02 & 3.18 & 4.38 & 0.13 & 2.73 & -2.58 & double scroll\\
1402 & 0.5993&-1.03 & 3.10 & 4.23 & 0.16 & 2.72 & -2.74 & periodic double scroll\\
1287 & 0.643&-1.07 & 2.61 & 3.36 & 0.26 & 2.61 & -3.55 & \\
\hline
\end{tabular}
\end{center}
\end{table}

\begin{figure}[htb]
\begin{minipage}[b]{0.47\linewidth}
\begin{center}
\epsfysize=120pt\epsfbox{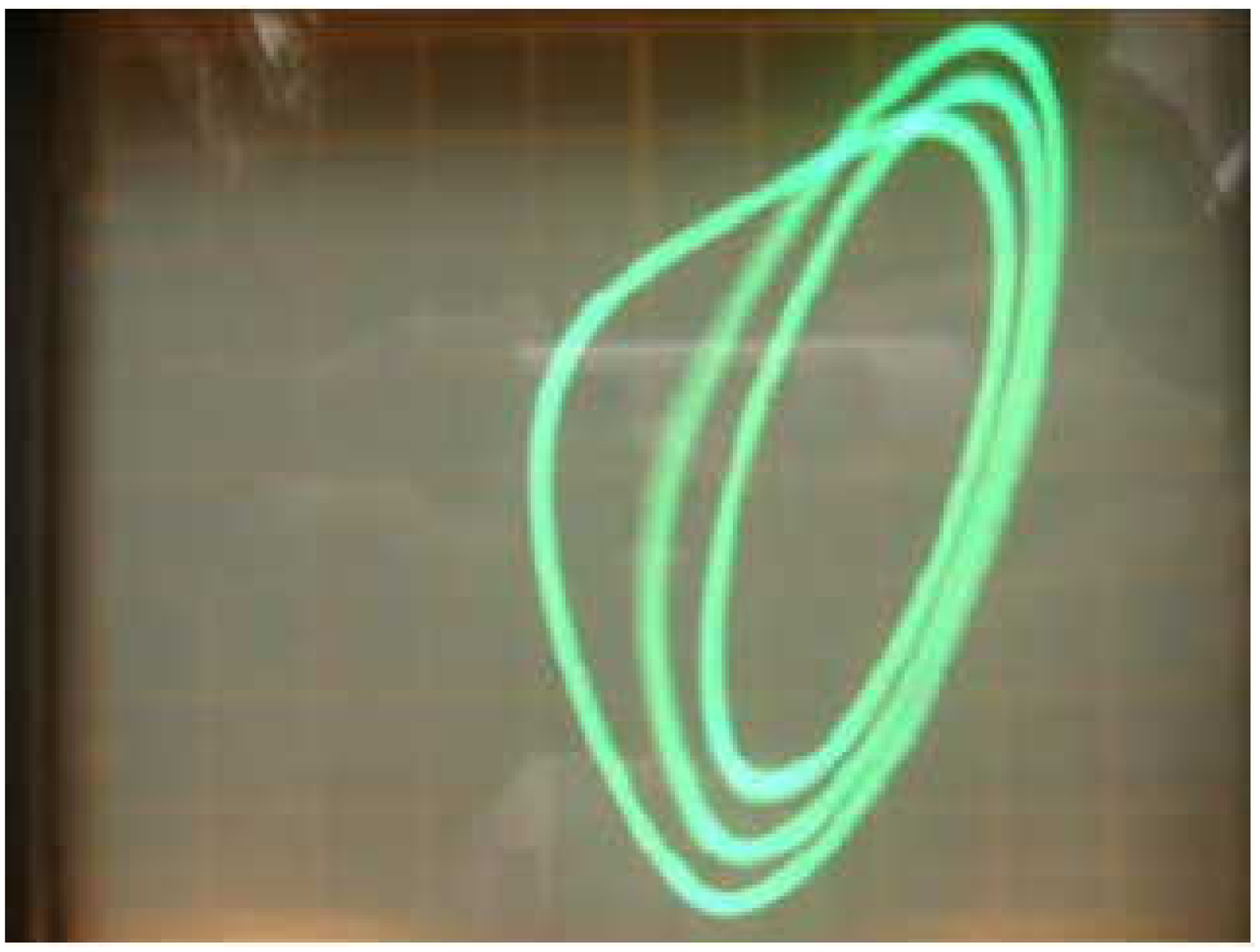}
\end{center}
\caption{The Chua's spiral impulse-1 type control.}\label{chua1_2s}
\end{minipage}
\hfill
\begin{minipage}[b]{0.47\linewidth}
\begin{center}
\epsfysize=120pt\epsfbox{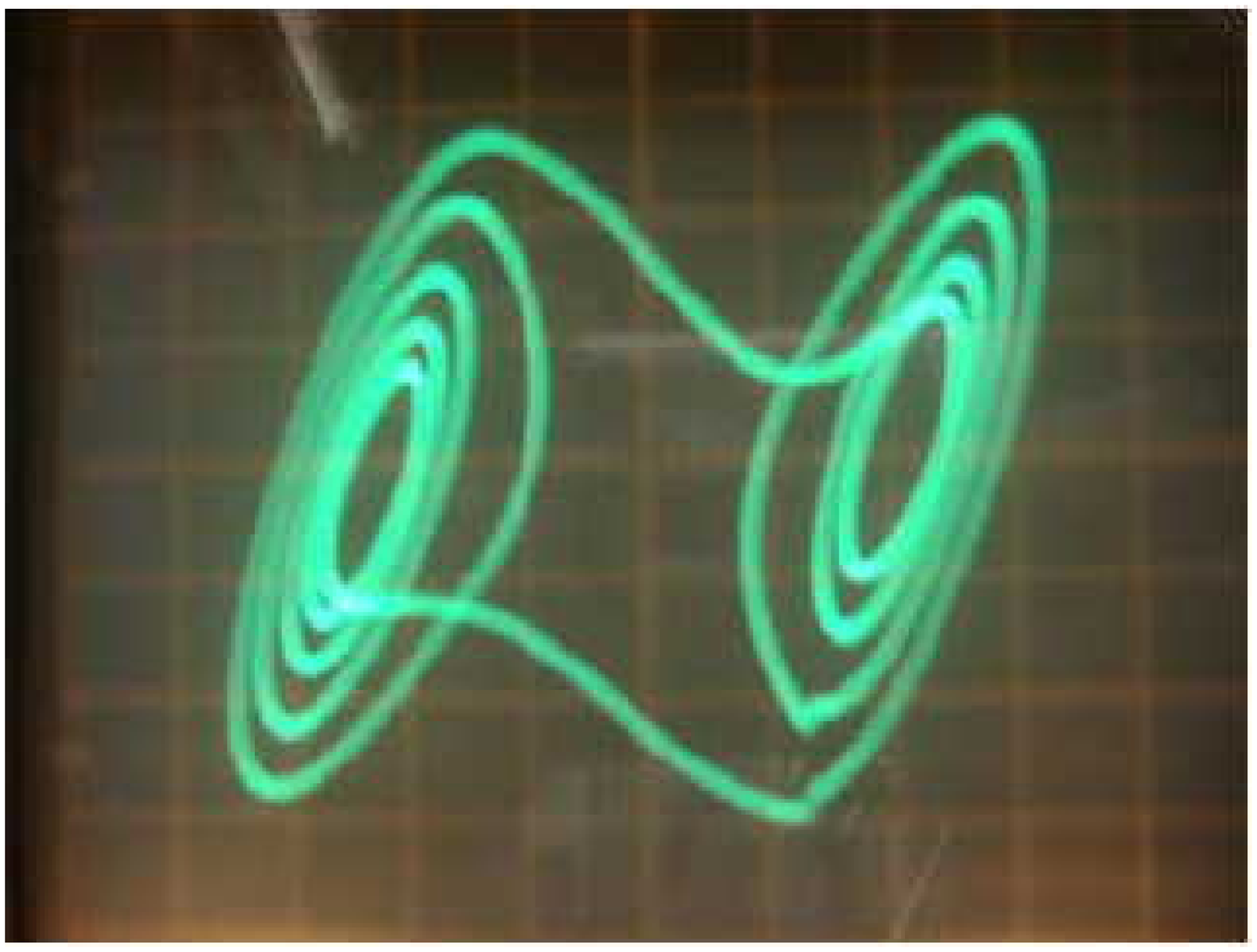}
\end{center}
\caption{The Chua's double scroll impulse-1 type control.}\label{impulse_1}
\end{minipage}
\end{figure}

\begin{figure}[htb]
\begin{minipage}[b]{0.47\linewidth}
\begin{center}
\epsfysize=120pt\epsfbox{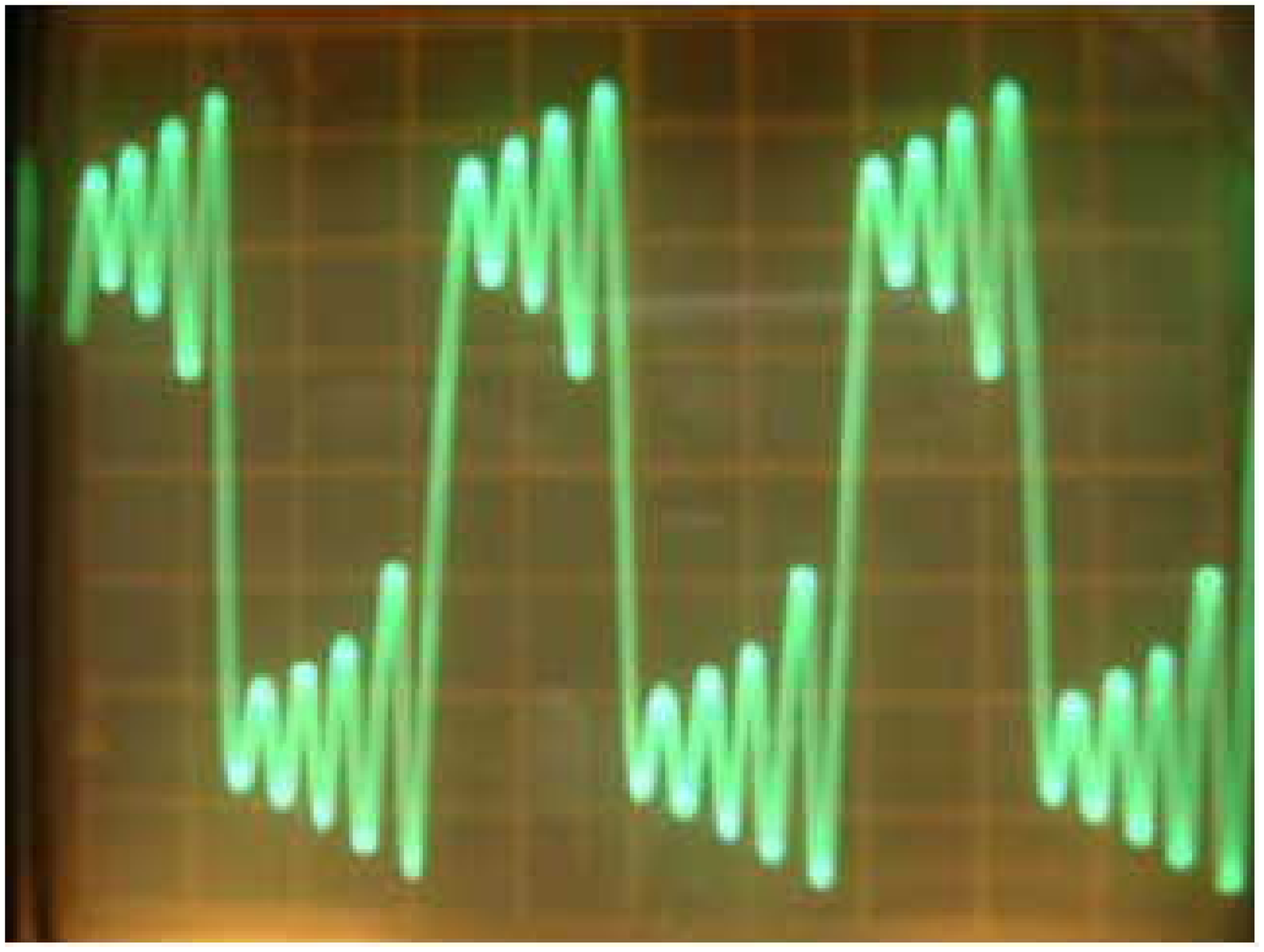}
\end{center}
\caption{The voltage of the capacitor-1 in double scroll impulse-1 type control.}\label{impulse_1x}
\end{minipage}
\hfill
\begin{minipage}[b]{0.47\linewidth}
\begin{center}
\epsfysize=120pt\epsfbox{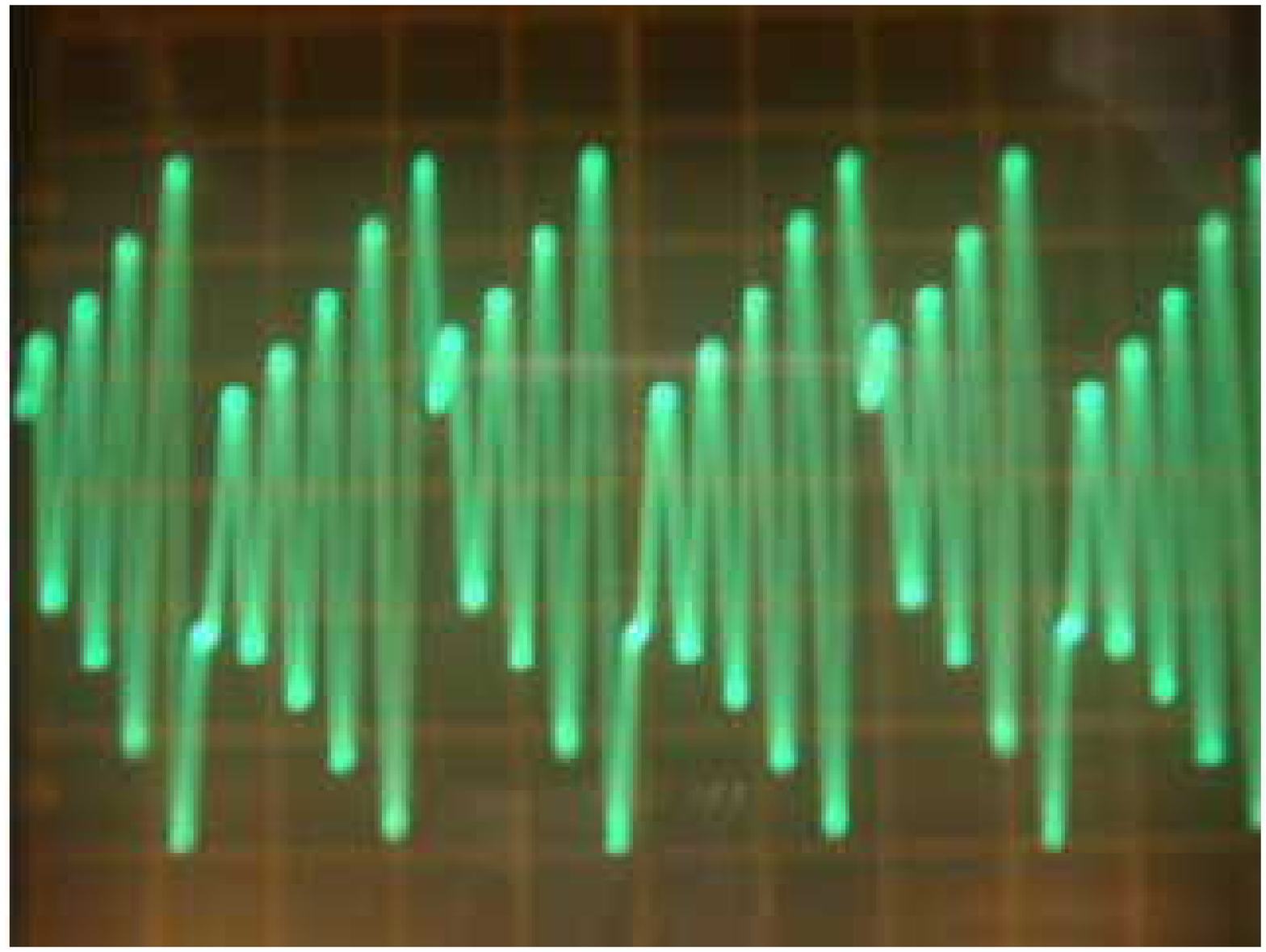}
\end{center}
\caption{The voltage of the capacitor-2 in double scroll impulse-1 type control.}\label{impulse_1y}
\end{minipage}
\end{figure}

The chaotic orbits can be guided to periodic orbits by triggering electric pulse as shown in Fig.\ref{chua1_2s} for the spiral and Fig.\ref{impulse_1} for the double scroll. 
In these photos, the abscissa is $v_{C1}$ and the ordinate is $v_{C2}$. The Figs.\ref{impulse_1x} and \ref{impulse_1y} are the time series of the $v_{C1}$ and the $v_{C2}$, respectively.

Since the electric field pulse applied to the system is not a delta function but contains a width, the shift $\delta y$ is not exactly equal to the height of the pulse. In the case of double scroll, the number of cycles in $z>0$ region and $z<0$ region depend on the voltage $v_{C2}$ of the sink that the orbit from unstable manifold is trapped, and so the exact quantitative simulation is impossible. But we can  study qualitative mechanism of the chaos control by simulation.

\section{Controlling the double scroll}

\begin{figure}[htb]
\begin{minipage}[b]{0.47\linewidth}
\begin{center}
\epsfysize=280pt\epsfbox{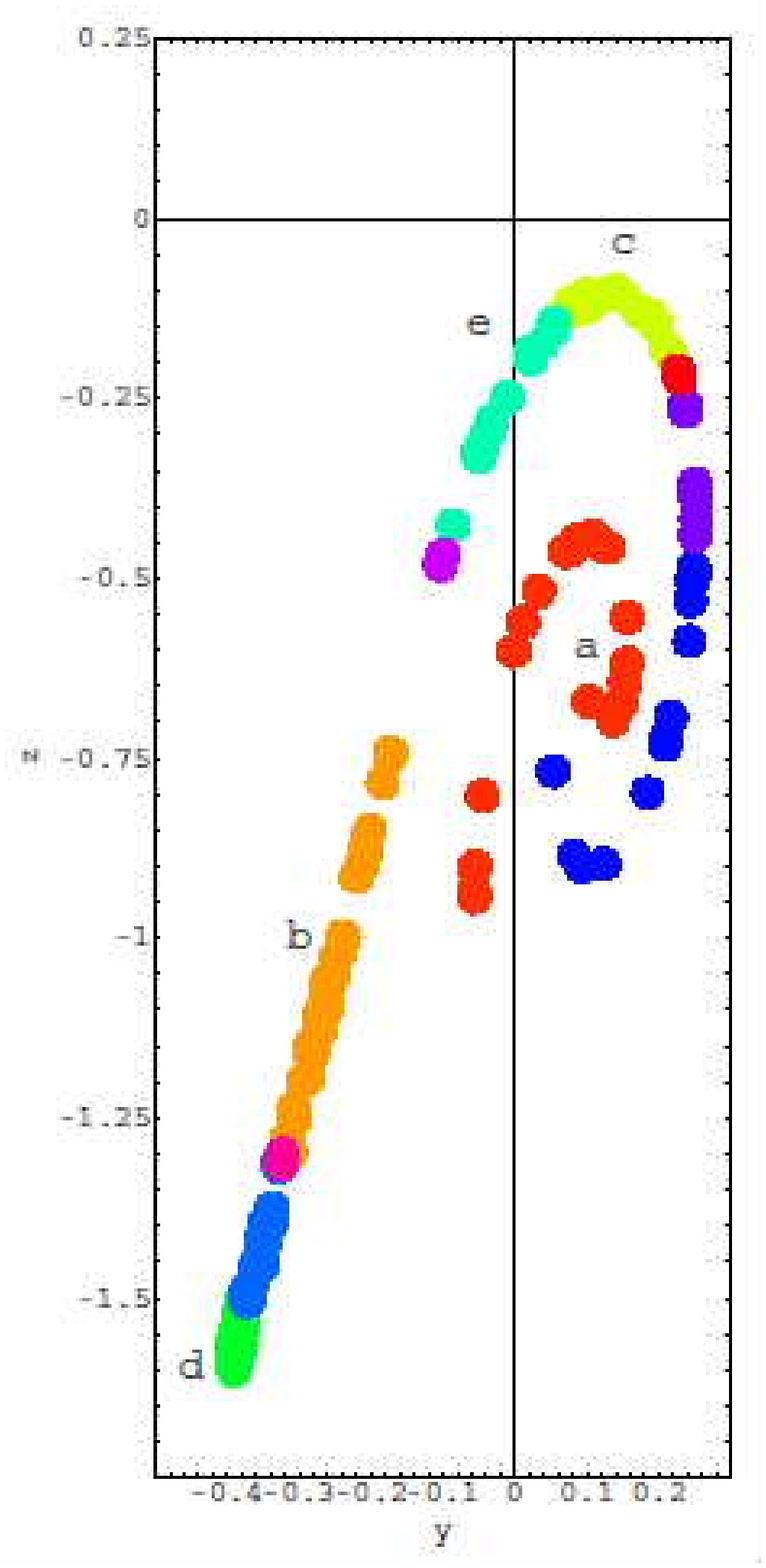}
\end{center}
\caption{The Poincar\'e map of the double scroll.}\label{dblescrl}
\end{minipage}
\hfill
\begin{minipage}[b]{0.47\linewidth}
\begin{center}
\epsfysize=280pt\epsfbox{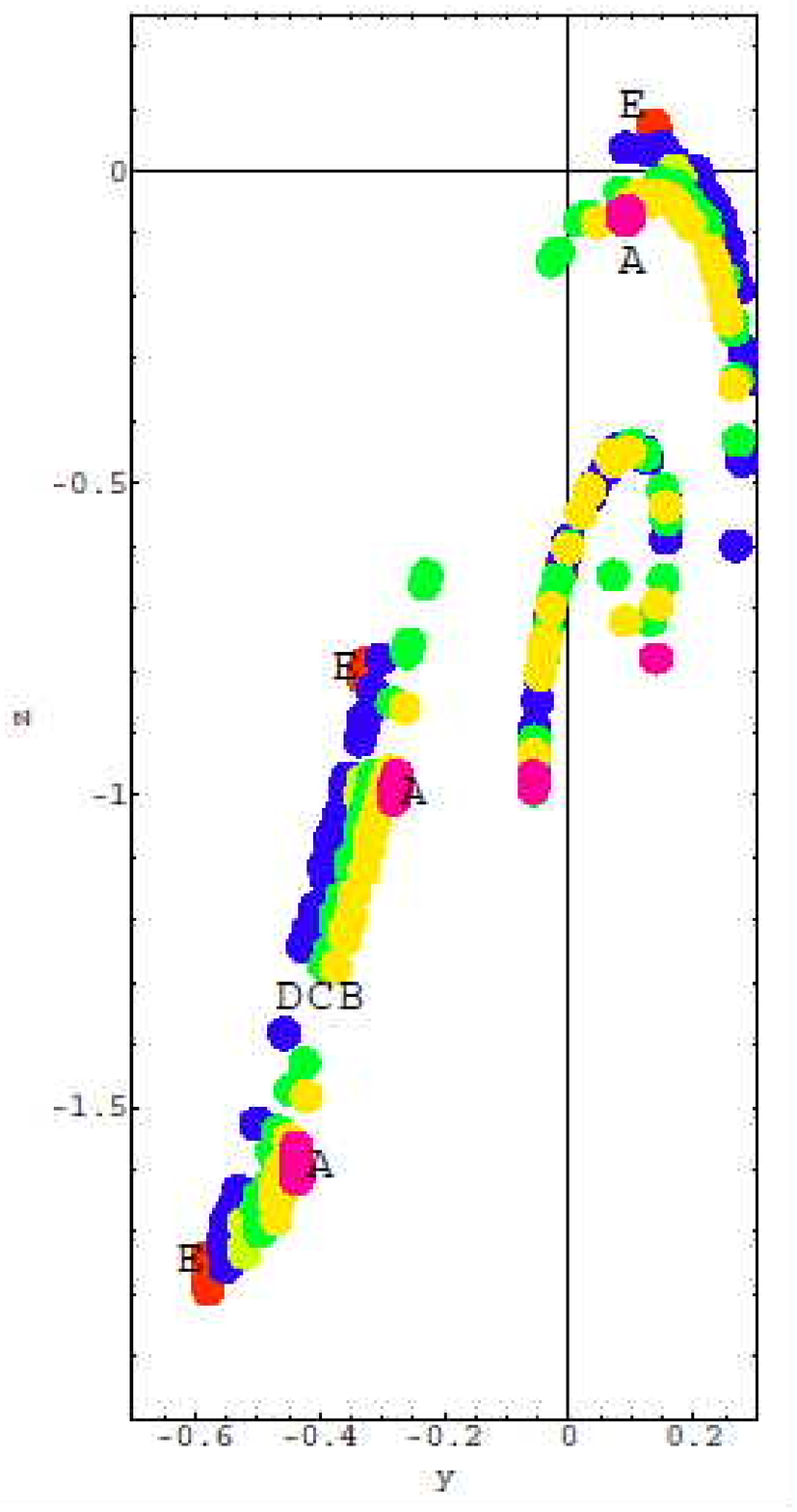}
\end{center}
\caption{The Poincar\'e map of the double scroll after triggering the impulse.}\label{dblescrl2}
\end{minipage}
\end{figure}

The Poincar\'e map on the plane of $x=1$ of the double scroll orbit obtained by using parameters after eq.(\ref{chua_eq})  is shown in Fig.\ref{dblescrl}. The points in the inner circle (region $a: -1<z<-0.4$) move to the outer branch (region $b: -0.3<y<-0.2, -1.3<z<-0.7$) and go to the top of the outer circle (region $c: 0.1<y<0.2$) and go to region $d$ or to its left neighbor (region $e: -0.1<y<0.1$) and go to complicated routes. When the trajectory goes to the latter region, the trajectory becomes complicated. Thus in order to realize periodic orbit, it is necessary to avoid the points trapped into region $e$.

The double scroll pattern of the Chua's circuit can be controlled by triggering  electric pulses to the system. We show two types of triggering, i.e. one pulse as the trajectory crosses $x=1$ (impulse-1) and pulses as the trajectory passes $x=1$ and $x=-1$ (impulse-2). 

When one triggers the impulse $\delta y$ when the orbit crosses $x=1$ from $x$ larger than 1, the curvature radius of the trajectory becomes larger. The points that the route takes for $\delta y=-0.041$ are points A, and the points for $\delta y=-0.06$ are the spiral that passes through B, the points for $\delta y=-0.08$ are the spiral that pases through C, and the points for $\delta y=-0.12$ are the spiral that passes through D. Isolated points E, near $(y,z)=(-0.6,-1.8)$, (-0.3,-0.7) and (0,-0.7) are for $\delta=-0.14$.  
The points that are trapped in $-0.1<y<0.1$ region become shifted to $0.1<y<0.2$ region, as shown in Fig.\ref{dblescrl2}. 

Using the above information, we trigger an electric impulse when the orbit passes the plane $x=1$, and study 1-d bifurcation diagram as a function of the height of the impulse. 

\begin{figure}[htb]
\begin{minipage}[b]{0.47\linewidth}
\begin{center}
\epsfysize=120pt\epsfbox{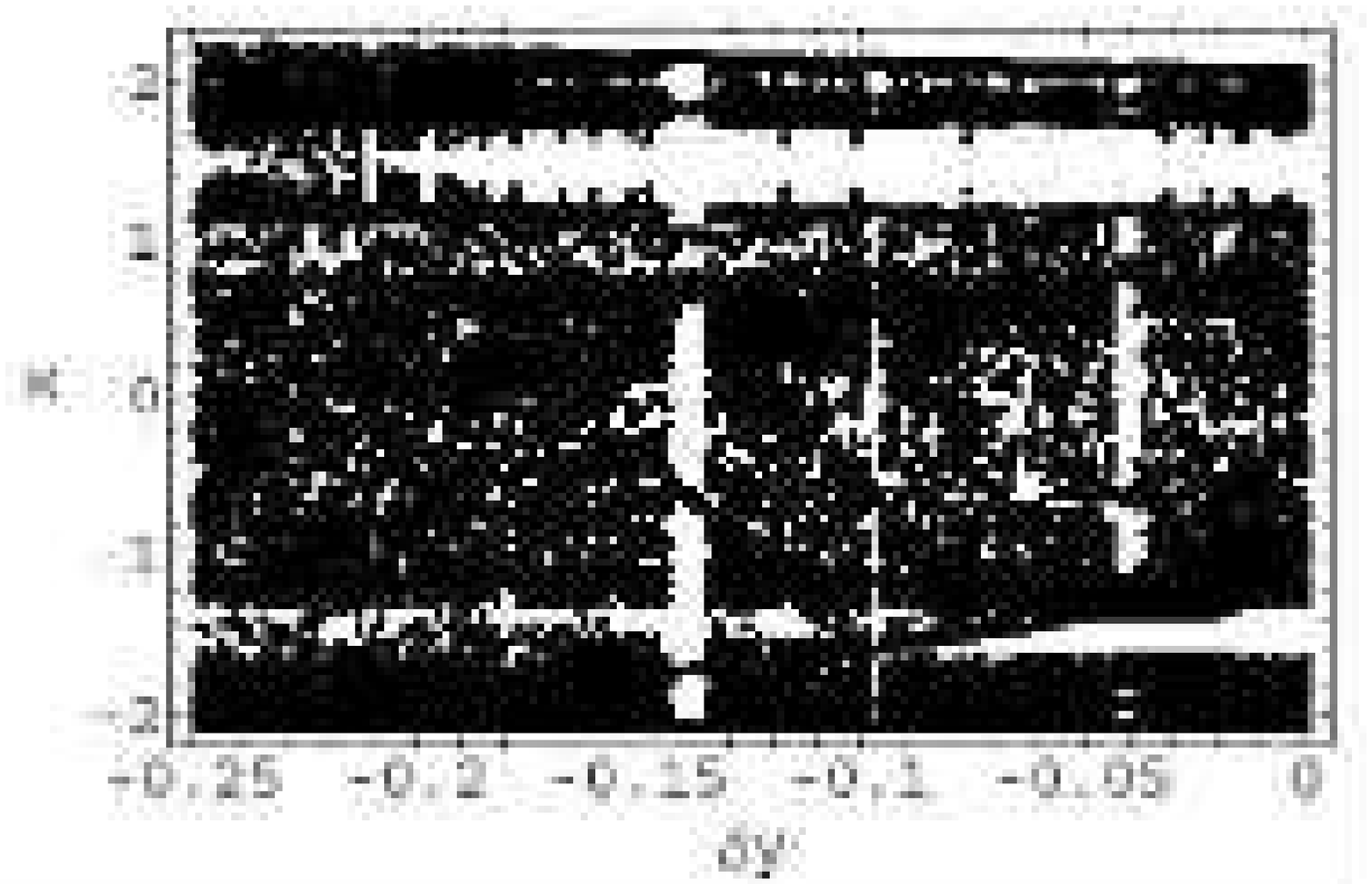}
\end{center}
\caption{The 1-d bifurcation diagram as a function of the impulse strength $\delta y$ of impulse-1 type.}\label{impulse_w}
\end{minipage}
\hfill
\begin{minipage}[b]{0.47\linewidth}
\begin{center}
\epsfysize=120pt\epsfbox{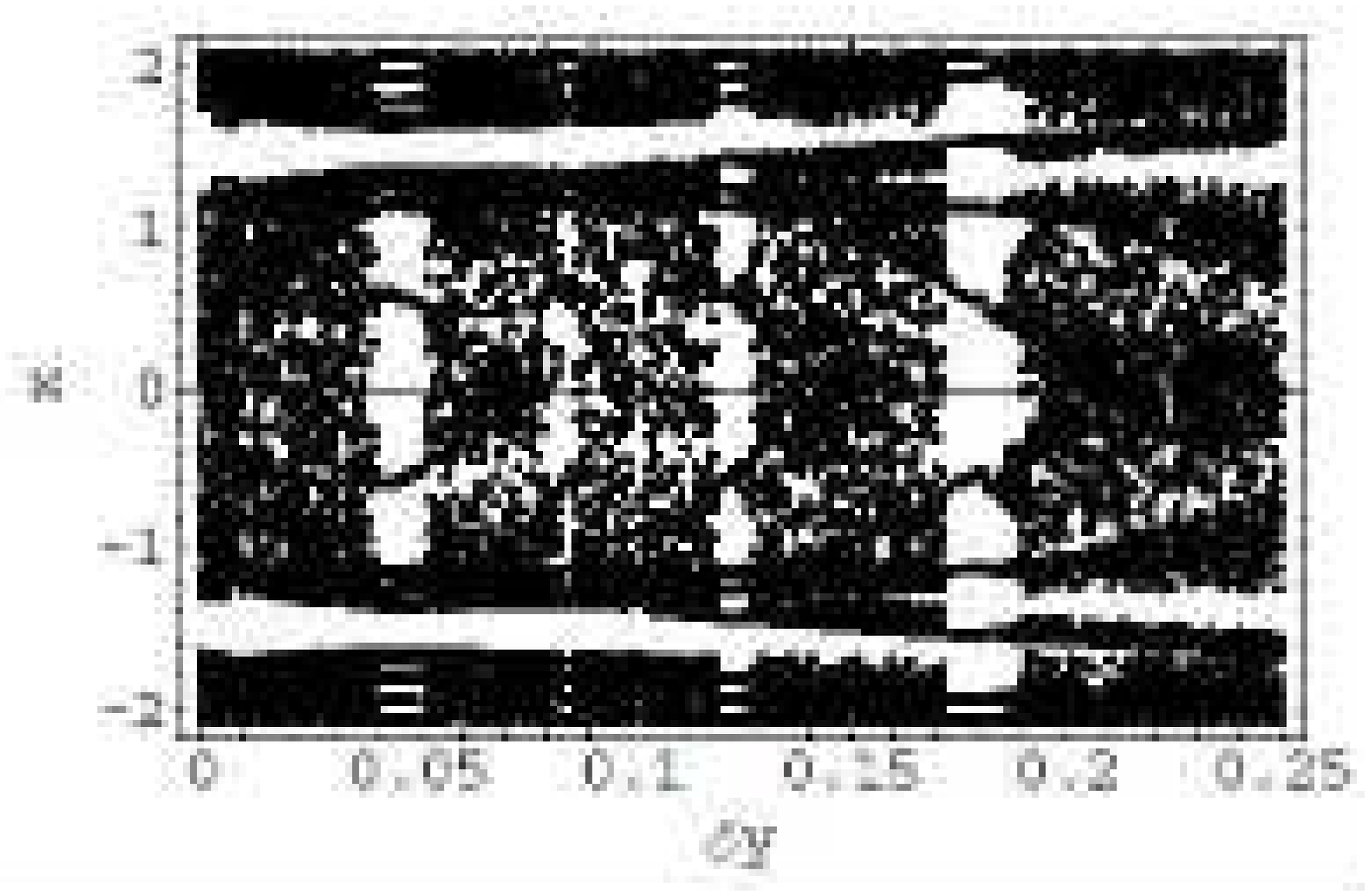}
\end{center}
\caption{The 1-d bifurcation diagram as a function of the impulse strength $\delta y$ of impulse-2 type.}\label{impulse2_w}
\end{minipage}
\end{figure}

We add a shift $\delta y$ as the trajectory passes the $x=1$ plane, and call it the impulse-1 type. In a plot of $x$ coordinate of the $y=0$ section as a function of the $\delta y$, we find windows for $\delta y=-0.41$, -0.09 and -0.14 as shown in Fig.\ref{impulse_w}. 
In the impulse-2 type, we add a shift $\delta y$ as the orbit passes the plane $x=1$ and $-\delta y$ as the orbit passes the plane $x=-1$. The windows of the impulse-2 type are shown in Fig.\ref{impulse2_w}.

The appearance of windows suggests a presence of heterocrinic orbits\cite{mc87}, or change of unstable orbits to stable orbits in a certain parameter region.

\section{The bifurcation analysis}
Chaotic system changes its Poincar\'e map when certain parameter of the dynamical equation is changed, and this phenomenon is known as bifurcation. In the present Chua's circuit, the parameters $\alpha$ and $\beta$ are fixed, but the strength of the additional electric pulse is a parameter that changes a complicated orbit to a torus. In order to analyze this qualitative changes,  we consider the 
 Poincar\'e mapping $T_\lambda$ i.e. a diffeomorphism from the state space ${\bf u}=(u_1,u_2,u_3)$ into ${\bf R}^3$
\begin{equation}
\Vec \phi=(\phi_1,\phi_2,\phi_3)=T_\lambda({\bf u}).
\end{equation}

As in \cite{kaw84}, we consider the first derivative of the Poincar\'e mapping
\begin{equation}
\frac{\partial}{\partial {\bf u}}T_\lambda({\bf u})=DT_\lambda({\bf u})=\frac{\partial{\boldmath \phi}}{\partial {\bf u}}(\tau,{\bf u},\lambda)
\end{equation}
which is a tensor
\begin{equation}
\frac{\partial {\Vec\phi}}{\partial {\bf u}}=\left(\begin{array}{ccc} 
   \frac{\partial \phi_1}{\partial u_1}& \frac{\partial \phi_1}{\partial u_2} & \frac{\partial \phi_1}{\partial u_3}\\
  \frac{\partial \phi_2}{\partial u_1} & \frac{\partial \phi_2}{\partial u_2} & \frac{\partial \phi_2}{\partial u_3}\\
   \frac{\partial \phi_3}{\partial u_1} & \frac{\partial \phi_3}{\partial u_2} &\frac{\partial \phi_3}{\partial u_3}\end{array}\right)
\end{equation}
that satisfies the differential eq.
\begin{equation}
\frac{d}{d\tau}\frac{\partial {\Vec\phi}}{\partial {\bf u}}=\frac{\partial {\bf f}}{\partial {\bf x}}\frac{\partial{\Vec\phi}}{\partial{\bf u}}\label{diffeq}
\end{equation}
where
\begin{equation}
\frac{\partial {\bf f}}{\partial {\bf x}}=\left(\begin{array}{ccc} 
   -s-s\frac{d g}{d x}& \alpha & 0\\
   1 & -1 & 1 \\
   0 & -\beta & 0\end{array}\right)
\quad{\rm and}\quad 
\frac{d g}{d x}=\left\{\begin{array}{cc} m_0 & x<-1\\
                                                      m_1& -1\leq x\leq 1\\
                                                      m_0& x>1\end{array}\right.
\end{equation}

The differential equation eq.(\ref{diffeq}) is solved by the 4th order Runge-Kutta method with initial condition $\displaystyle \frac{\partial{\Vec\phi}}{\partial{\bf u}}=I$ at initial $\tau$ to $\tau+T_{per}$, where $\tau$ is chosen at the point when the pulse in $y$ direction is triggered and $T_{per}$ is the period between the pulses are triggered. We use data ${\bf u}(\tau)$ of the periodic trajectory for fixing the time that the pulse is triggered and thus the $y$ coordinate is shifted.  We measure eigenvalues of $\displaystyle\frac{\partial {\Vec\phi}}{\partial {\bf u}}$ at each $\tau$ and store.

In most cases, after a few hundred steps, one of the three eigenvalues of the first derivative of the Poincar\'e mapping damps to very close to 0 and the diffeomorphism defined by $\displaystyle \frac{\partial{\bf f}}{\partial {\bf x}}$ becomes essentially 2-dimensional.

\section{Impulse control of the double scroll}
We simulate the Shil'nikov's chaos by choosing the initial condition
${\bf u}(1)=(0.01,0.01.0.01)$ and take $d\tau=0.01$ and make 50000 steps.
Using the parameter $\alpha=9, \beta=14.28$ we obtained the double scroll Fig. \ref{chua1_2d}.

We consider 2 types of controlling Shil'nikov's double scroll by impulse.
\begin{itemize}
\item Impulse-1a type: When $x$ passes 1 from $x_{n}>x_{n+1}$ shift $y$ to $y-0.041$.

\item Impulse-1b type: When $x$ passes 1 from $x_{n}>x_{n+1}$ shift $y$ to $y-0.097$.

\item Impulse-1c type: When $x$ passes 1 from $x_{n}>x_{n+1}$ shift $y$ to $y-0.147$.

\item Impulse-2a type:

 a) When $x$ passes $1$ from $x_{n}>x_{n+1}$ and $y\lesssim -0.37$ shift $y$ to $y-0.05$

b) When $x$ passes $-1$ from $x_{n}<x_{n+1}$ and $y\gtrsim 0.37$ shift $y$ to $y+0.05$

\item Impulse-2b type:

 a) When $x$ passes $1$ from $x_{n}>x_{n+1}$ and $y\lesssim -0.37$ shift $y$ to $y-0.18$

b) When $x$ passes $-1$ from $x_{n}<x_{n+1}$ and $y\gtrsim 0.37$ shift $y$ to $y+0.18$
\end{itemize}

\subsection{The impulse-1a type control}
\begin{figure}[htb]
\begin{minipage}[b]{0.47\linewidth}
\begin{center}
\epsfysize=120pt\epsfbox{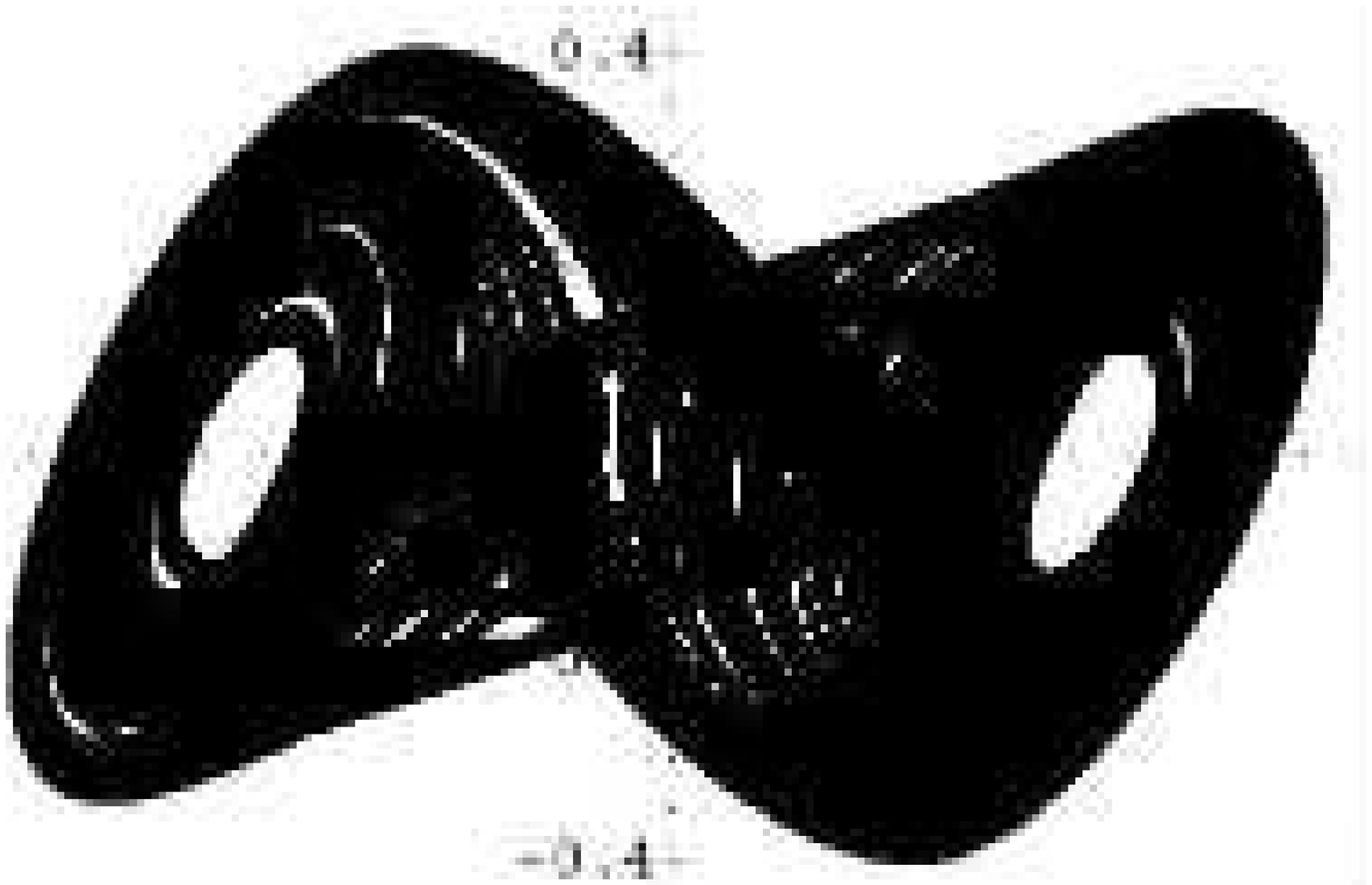}
\end{center}
\caption{The trajectory of the double scroll without impulse.}\label{chua1_2d}
\end{minipage}
\hfill
\begin{minipage}[b]{0.47\linewidth}
\begin{center}
\epsfysize=120pt\epsfbox{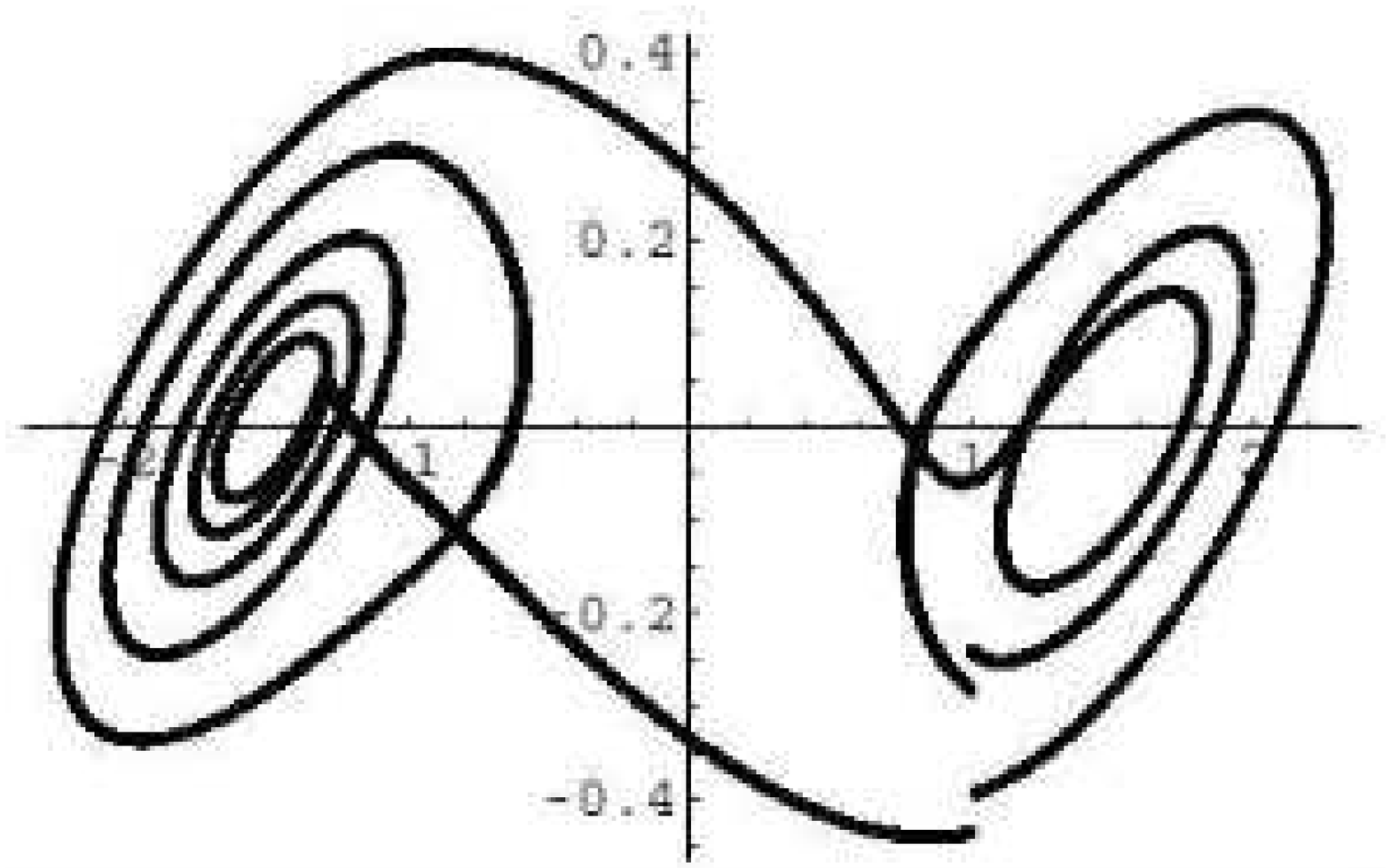}
\end{center}
\caption{The trajectory of the double scroll controlled by impulse-1a type.}\label{impulse_1a}
\end{minipage}
\end{figure}

The simulation results of the controlled trajectories (the last 5000 steps of the 50000 steps) by the impulse-1a type is shown in Fig. \ref{impulse_1a}.

\begin{figure}[htb]
\begin{minipage}[b]{0.47\linewidth}
\begin{center}
\epsfysize=120pt\epsfbox{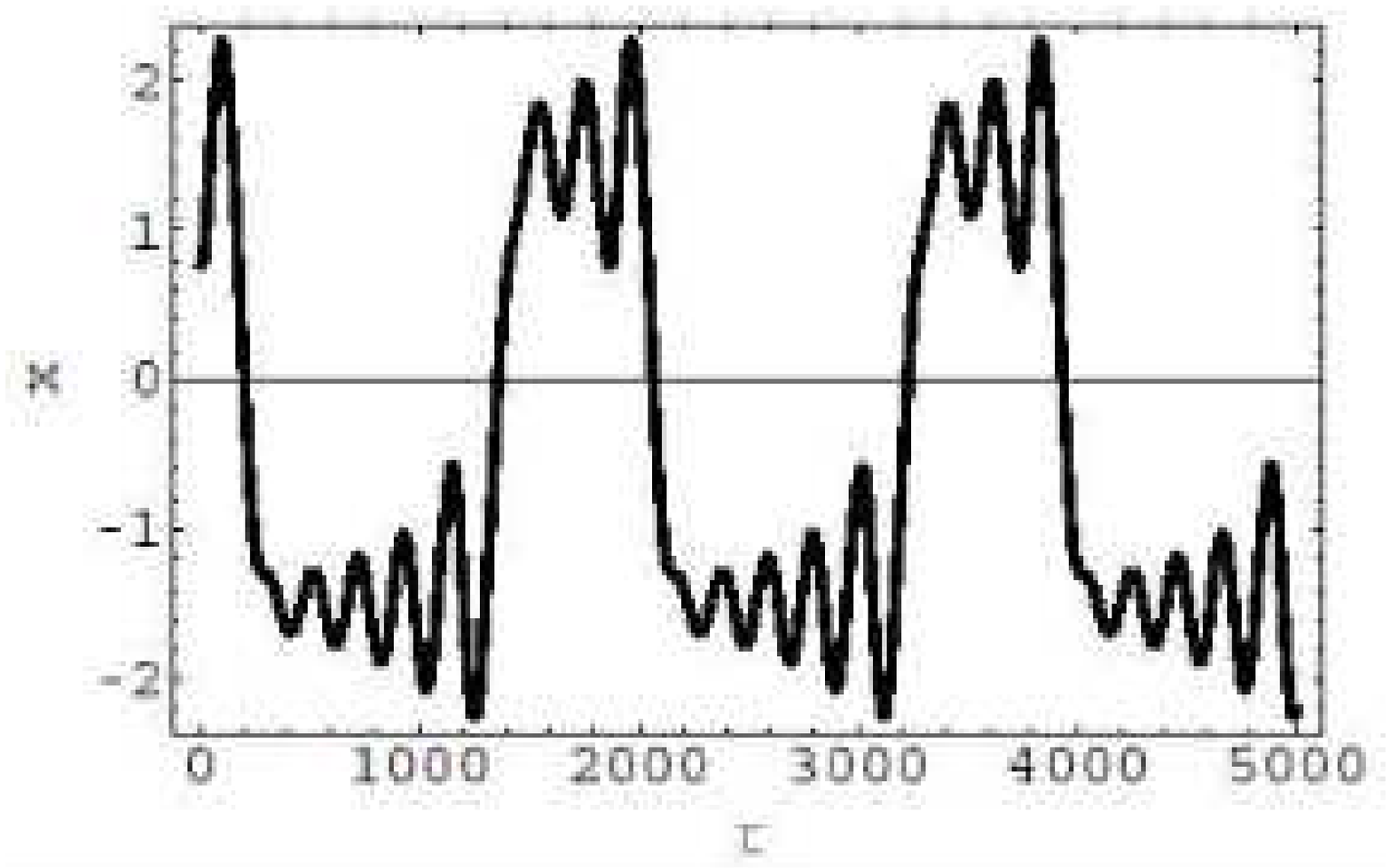}
\end{center}
\caption{The $x$ coordinate of the trajectory of impulse-1a type.}\label{impulse_1xn}
\end{minipage}
\hfill
\begin{minipage}[b]{0.47\linewidth}
\begin{center}
\epsfysize=120pt\epsfbox{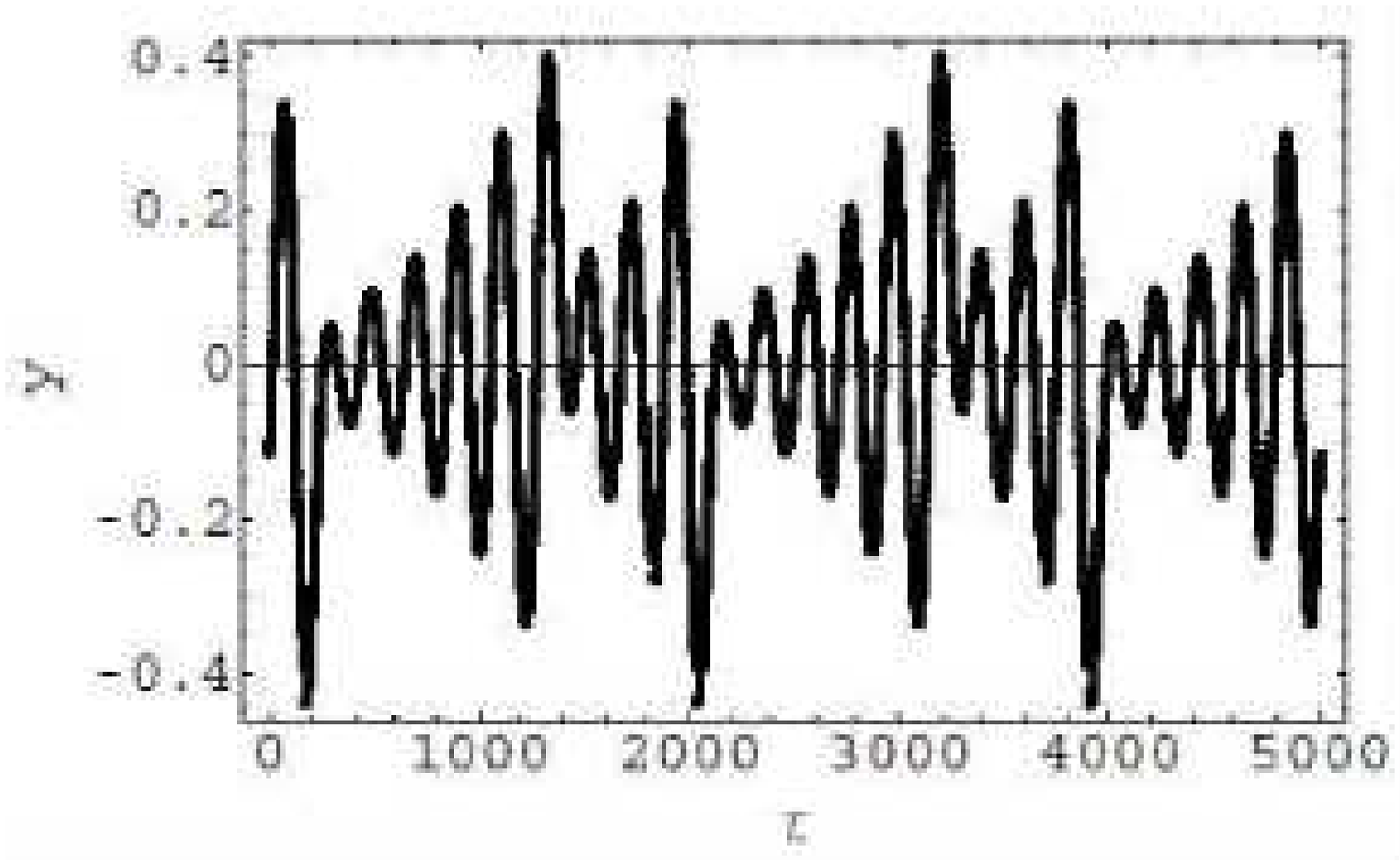}
\end{center}
\caption{The imaginary part(smooth curve that crosses 1) and real part(bending curve) of the complex eigenvalue and the small real eigenvalue of impulse-1a type.}\label{impulse_1mu}
\end{minipage}
\end{figure}

The $x$  coordinate of the trajectory obtained by applying the shift $\delta y=-0.041$ are shown in Figs.\ref{impulse_1xn} .

\begin{figure}[htb]
\begin{minipage}[b]{0.47\linewidth}
\begin{center}
\epsfysize=120pt\epsfbox{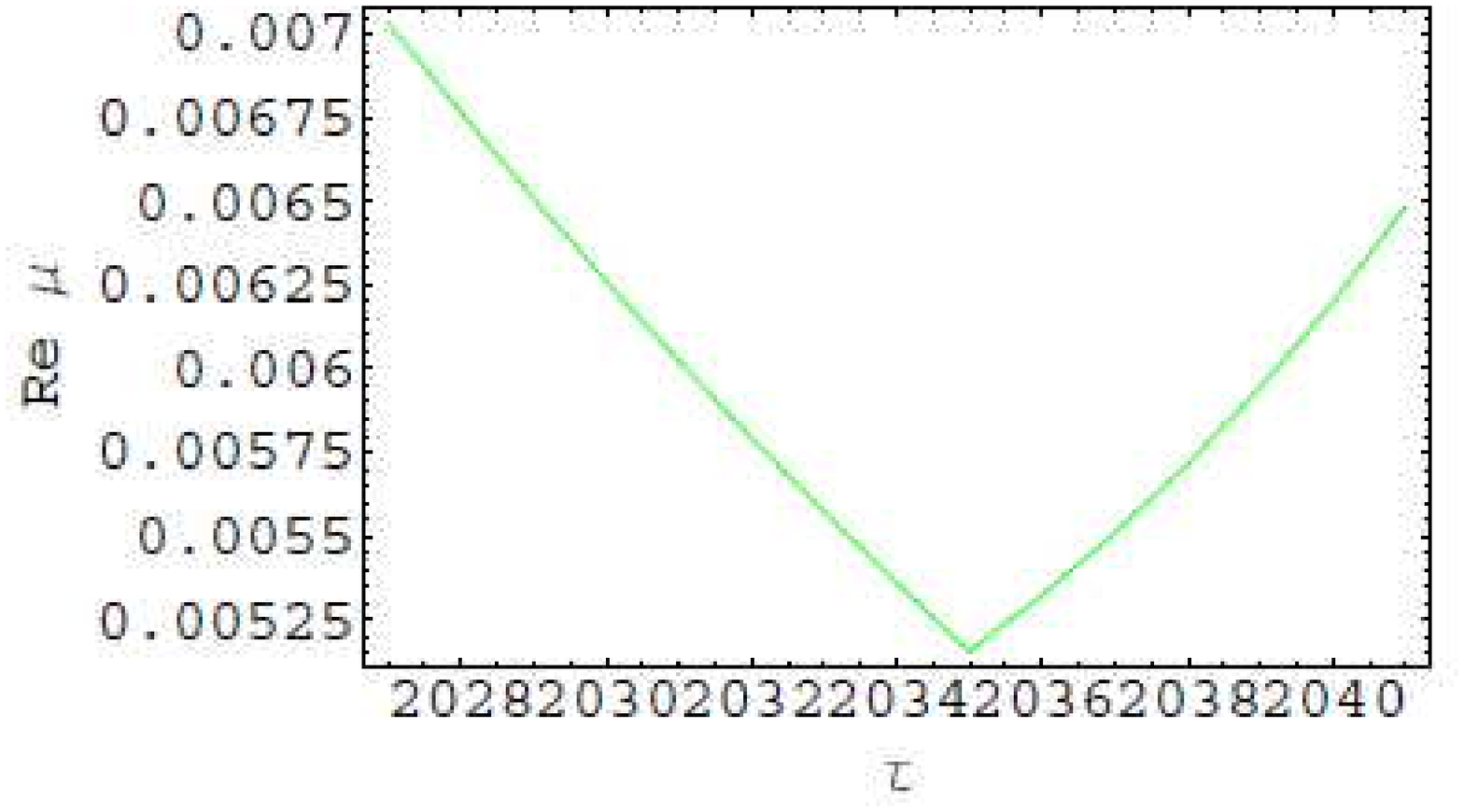}
\end{center}
\caption{Magnification of the smallest eigenvalue of the first derivative of the Poincar\'e mapping of the impulse-1a type.}\label{niiya1_3}
\end{minipage}
\hfill
\begin{minipage}[b]{0.47\linewidth}
\begin{center}
\epsfysize=120pt\epsfbox{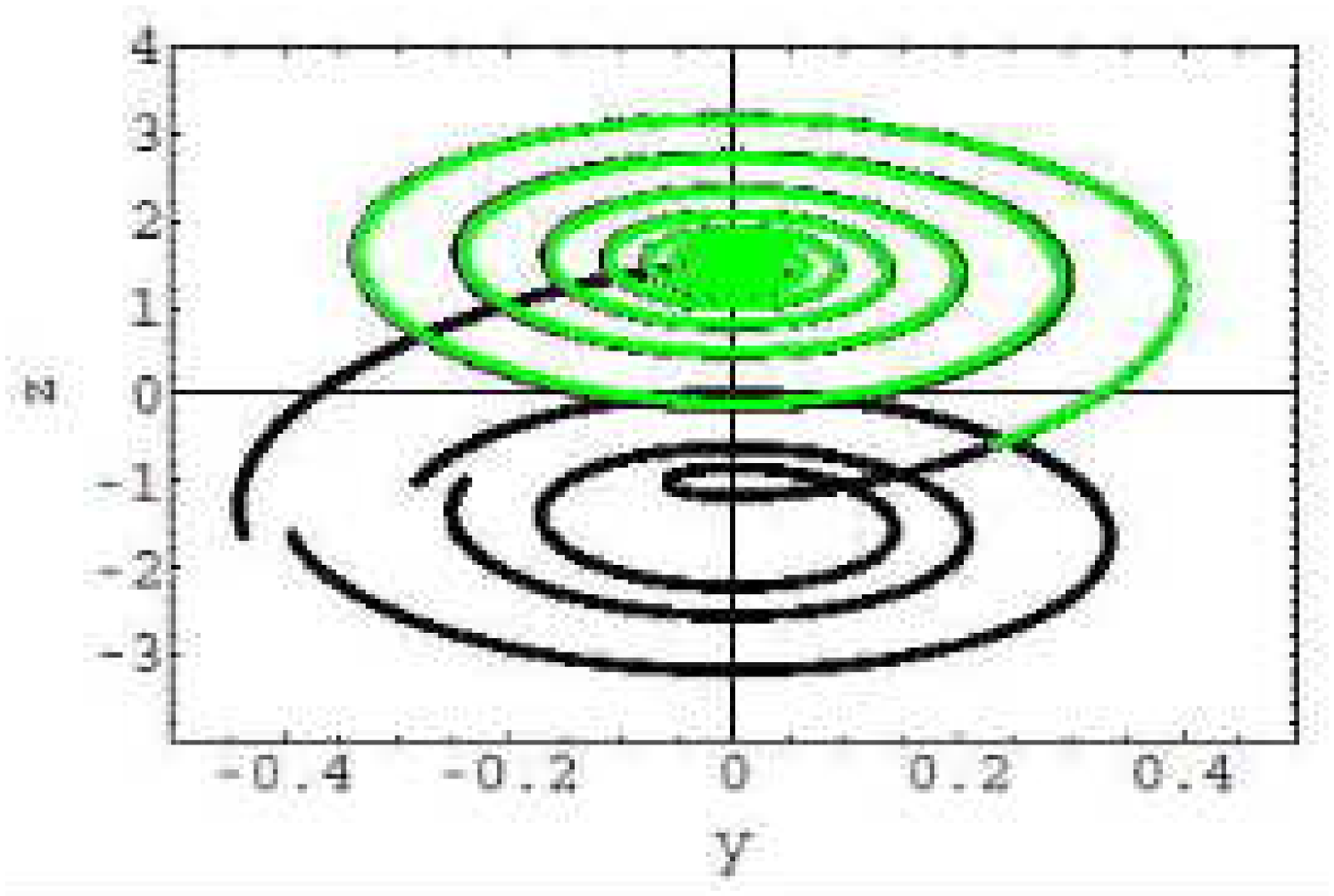}
\end{center}
\caption{Tangency of the controlled orbit (impulse-1a) and the homoclinic orbit.}\label{newhouse1}
\end{minipage}
\end{figure}

The first derivative of the Poincar\'e mapping is calculated from $\tau=1838$ when the $\delta y$ is applied to $\tau=3705$ when the subsequent  $\delta y$ is applied.  Its eigenvalues consist of a pair of complex and one real. 
The behaviors of the real eigenvalue and two complex eigenvalues  before and after the second pulse are shown in Fig.\ref{impulse_1mu} and in \ref{niiya1_3}.  The imaginary part of the complex eigenvalues crosses Im$\mu=\pm 1$ near $\tau$ when the impulse is applied. The real part of the complex eigenvalue is positive but the slope of increasing tendency is reduced when the impulse is applied.

The trajectory of the impulse-1a type suggests that the orbit in the unstable manifold $W^u$ is perturbed in the region $\Lambda_2$ and shifted to the stable manifold $W^s$ and  near the fixed point $P_+=(-1.5,0,1.5)$ crossing to the unstable manifold $W^u$ occurs in the region $\Lambda_1$. 
In Fig.\ref{newhouse1}, we show the homoclinic orbit started from the vicinity of the fixed point $P_+'=(-1.5+0.001,0,1.5)$ and the controlled trajectory of the impulse-1a type. The figure suggests that after the impulse, the trajectory is absorbed into the homoclinic orbit and the tangency of the unstable manifold $W^u(\Lambda_-)$ and the stable manifold $W^s(\Lambda_+)$ occurs, where $\Lambda_\pm$ implies the region near the fixed point $P_\pm$, respectively. 
Products of the three eigenvalues after $\tau=2035$ are almost constant of 0.0069(1), which implies that the orbit is on $W^s(\Lambda_+)$ 
It is interesting that the absolute value of the imaginary part of the eigenvalue becomes smaller than 1 as the trajectory shifts to the manifold $W^s(\Lambda_+)$.

\subsection{The impulse-1b type control}
\begin{figure}[htb]
\begin{minipage}[b]{0.47\linewidth}
\begin{center}
\epsfysize=120pt\epsfbox{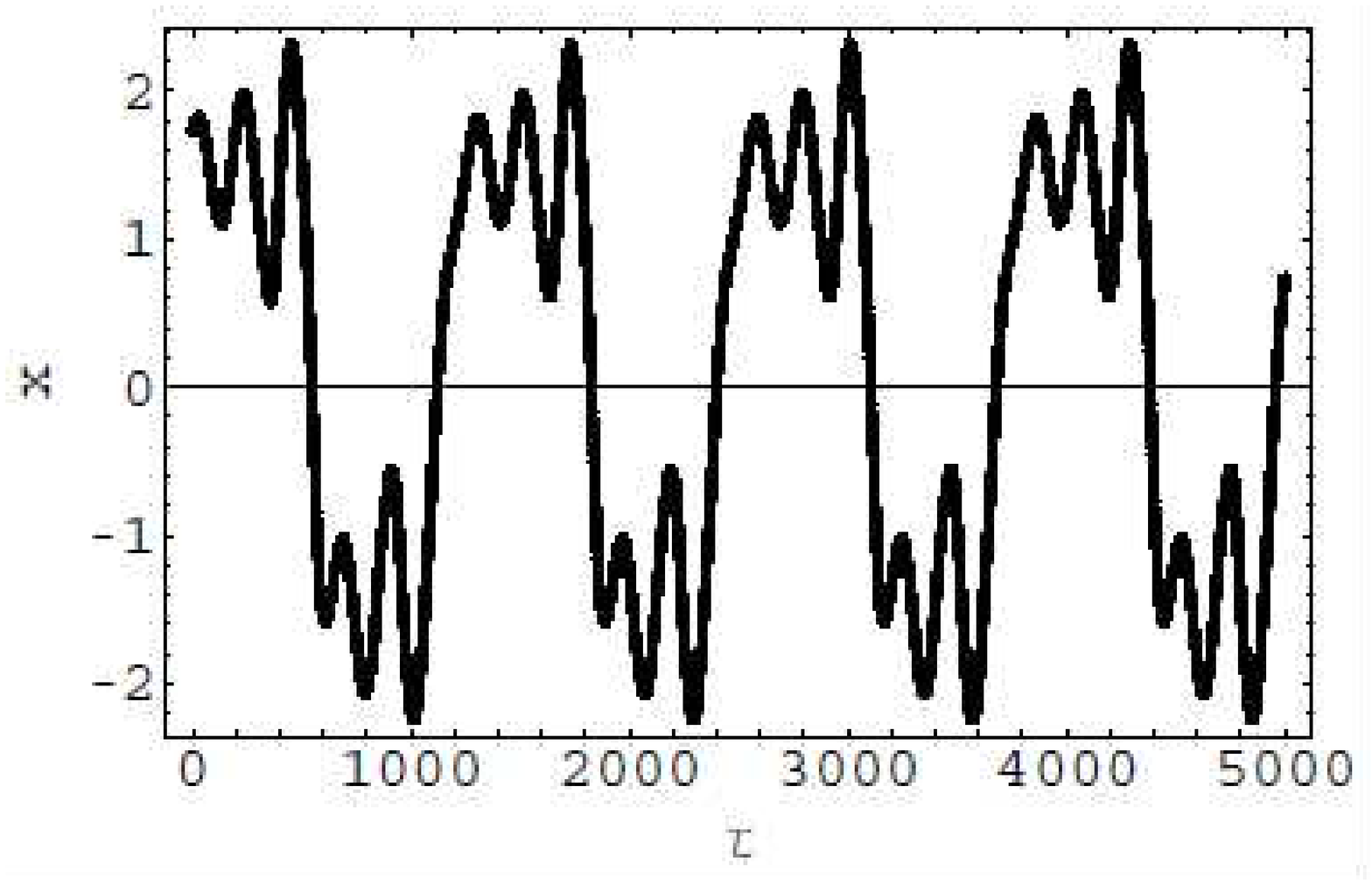}
\end{center}
\caption{The $x$ coordinate of the trajectory of impulse-1b type.}\label{impulse_6x}
\end{minipage}
\hfill
\begin{minipage}[b]{0.47\linewidth}
\begin{center}
\epsfysize=120pt\epsfbox{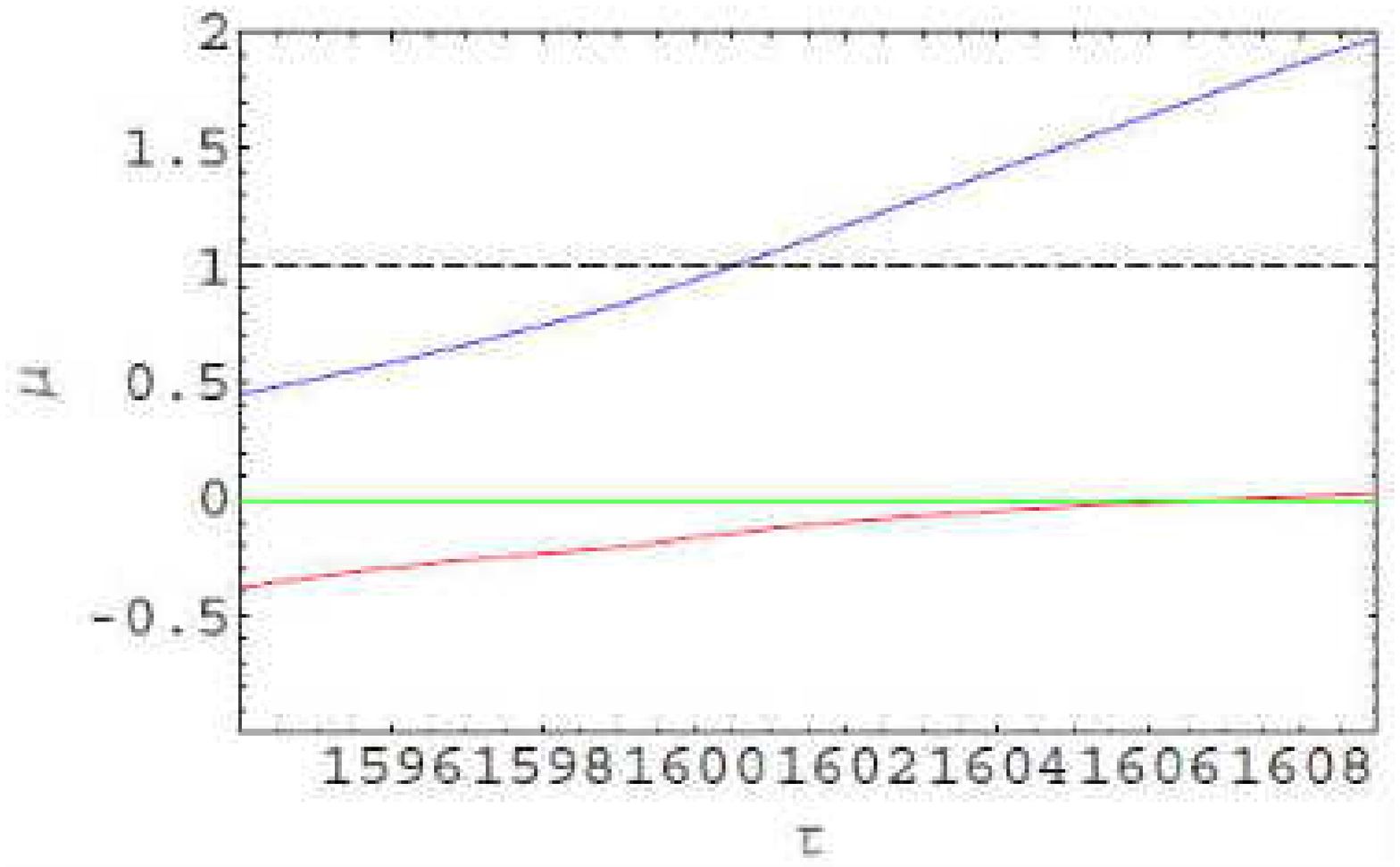}
\end{center}
\caption{The real part of the eigenvalue of the first derivative of the Poincar\'e mapping of the impulse-1b type.}\label{impulse_6mu}
\end{minipage}
\end{figure}

The second window in the 1-d bifurcation diagram exists near $\delta y=-0.097$.
The $x$ coordinate of the trajectory for this perturbation is shown in  Fig.\ref{impulse_6x}. There are  3 cycle oscillation in $y>0$ and 3 cycle oscillation in $y<0$.
\begin{figure}[htb]
\begin{center}
\epsfysize=120pt\epsfbox{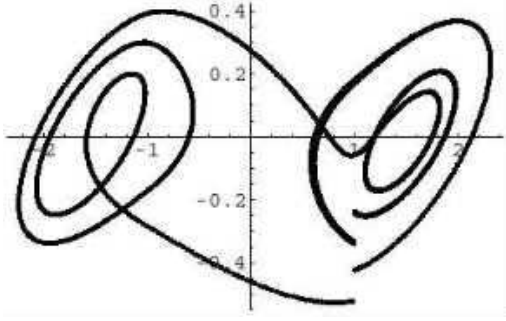}
\end{center}
\caption{The trajectory of the Chua's circuit impulse-1b type.}\label{impulse_6}
\end{figure}

The behavior of the eigenvalues of the first derivative of the Poincar\'e mapping in the case of $\delta y=-0.097$ is different from that of $-0.041$. We measure it from $\tau=319$ when the first $\delta y$ is applied, to $\tau=1599$ when the subsequent $\delta y$ is applied.
In Fig.\ref{impulse_6mu}, real eigenvalues of the first derivative of the Poincar\'e mapping near the time when the second pulse is applied are shown. A crossing of the eigenvalue with the line of $\mu=1$ occurs as the pulse is triggered. One of the three eigenvalues of the first derivative of the Poincar\'e mapping is zero and the product of the rest two eigenvalues before $\tau=1599$ is constant (-0.170(2)), which means that the orbit has been on the $W^s(\Lambda_+)$.

\subsection{The impulse1-c type control}
\begin{figure}[htb]
\begin{minipage}[b]{0.47\linewidth}
\begin{center}
\epsfysize=120pt\epsfbox{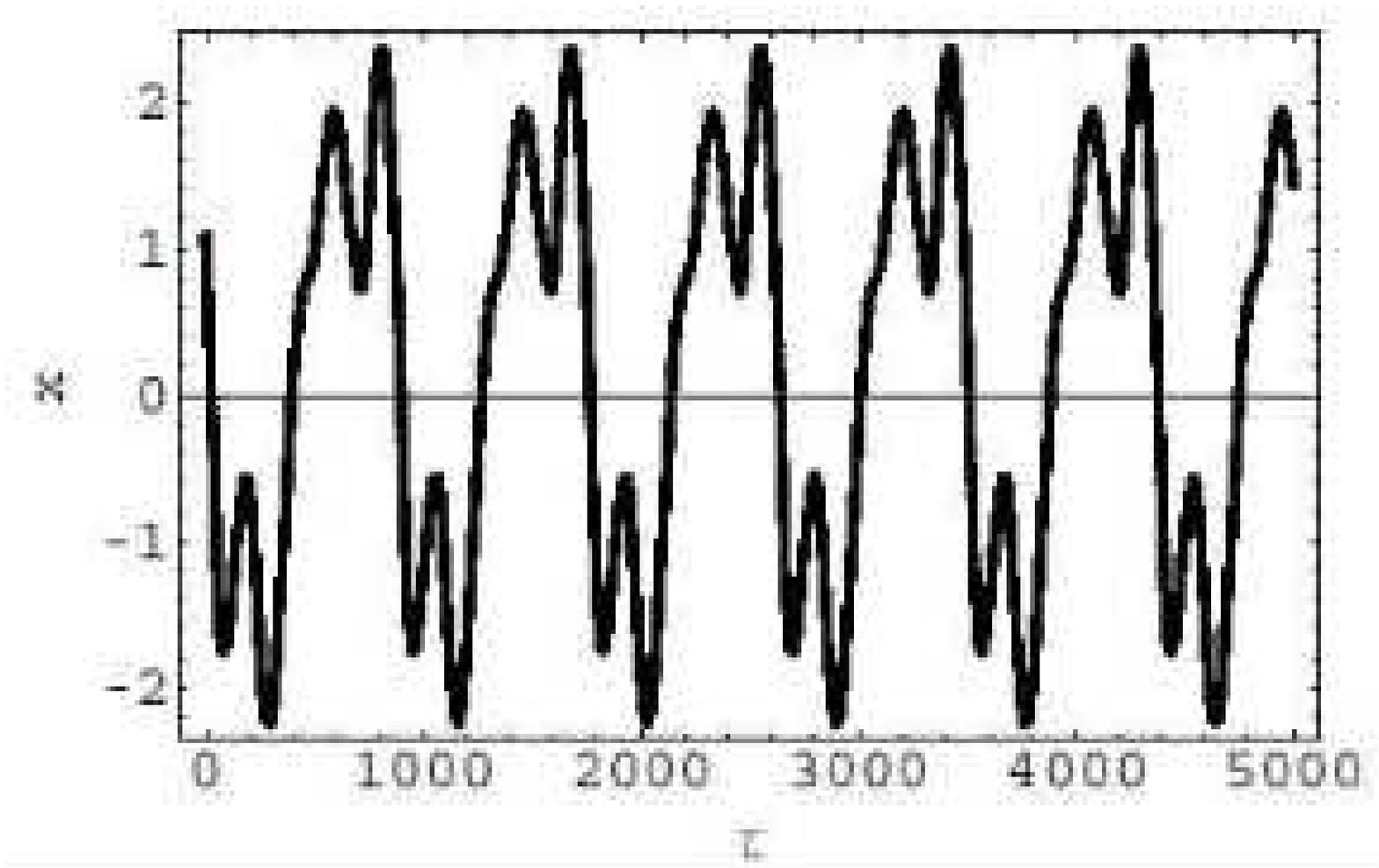}
\end{center}
\caption{The $x$ coordinate of the trajectory of impulse-1c type.}\label{impulse_5x}
\end{minipage}
\hfill
\begin{minipage}[b]{0.47\linewidth}
\begin{center}
\epsfysize=120pt\epsfbox{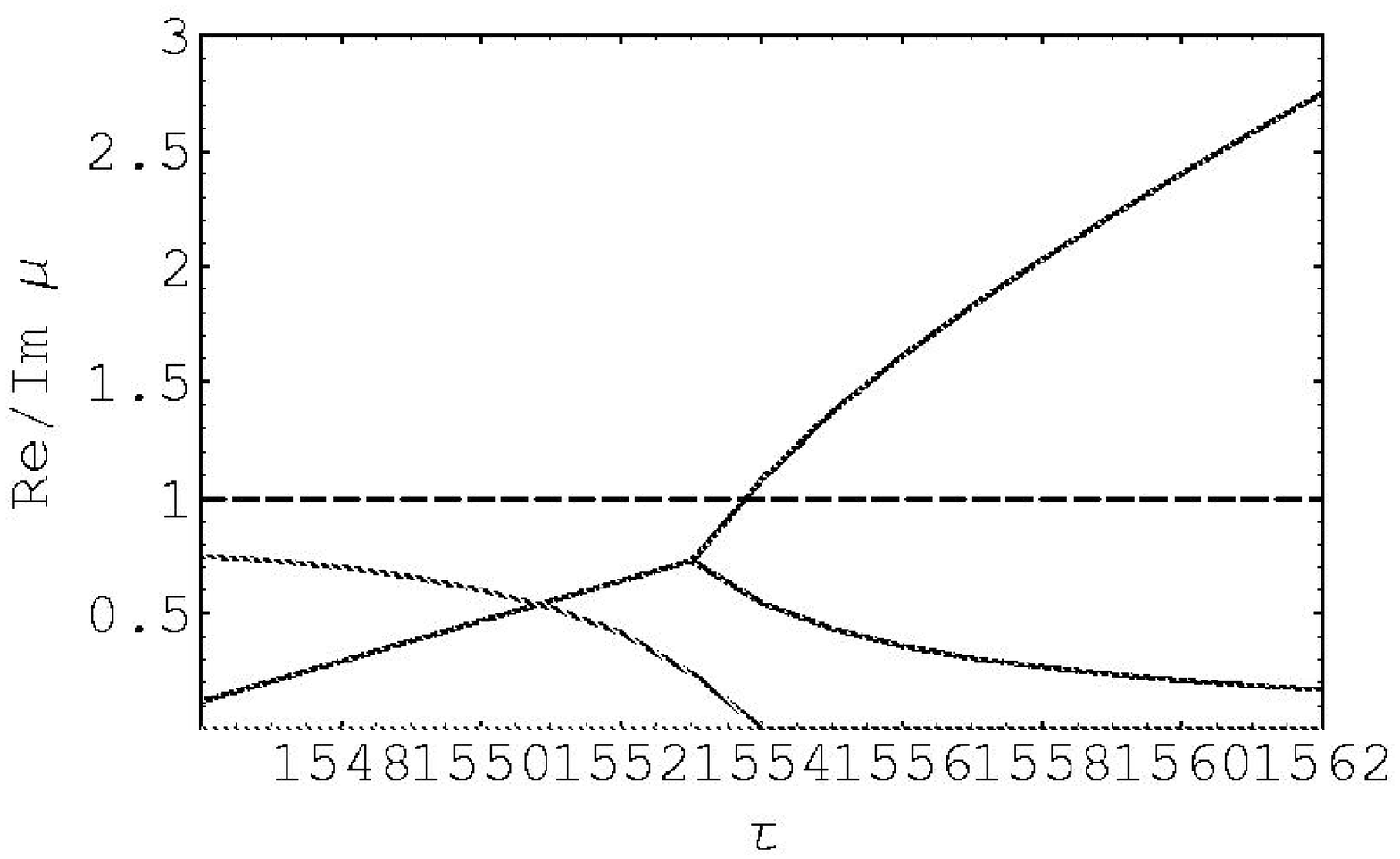}
\end{center}
\caption{The real part ( branching at $\tau=1554$) and the imaginary part (vanishing at $\tau=1554$) of the eigenvalue of the first derivative of the Poincar\'e mapping of impulse-1c type.}\label{impulse_5r}
\end{minipage}
\end{figure}

Corresponding to the relatively large window in the 1-d bifurcation diagram near $\delta y=-0.14$, we trigger $\delta y=-0.14$ in the impulse-1c type. There are  2 cycle oscillation in $y>0$ and 2 cycle oscillation in $y<0$.
\begin{figure}[htb]
\begin{center}
\epsfysize=160pt\epsfbox{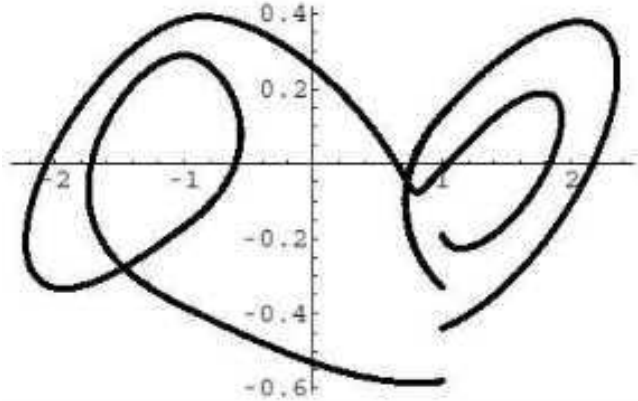}
\end{center}
\caption{The trajectory of the Chua's circuit impulse-1c type.}\label{impulse_5}
\end{figure}
The $x$ coordinate of the trajectory for impulse-1c are shown in  Fig. \ref{impulse_5x}.

The first derivative of the Poincar\'e mapping is calculated from $\tau=685$ where the $\delta y$ is applied to $\tau=1609$ where the subsequent $\delta y$ is applied.  In Figs.\ref{impulse_5r}, the eigenvalues are shown. Before the impulse, there is a pair of complex eigenvalues and after the impulse, a pair of real eigenvalues appear. Product of the three eigenvalues is constant (0.58(1)) before $\tau=1554$ and the orbit has been on $W^s(\Lambda_+)$. After changing from complex eigenvalue to real eigenvalue, one eigenvalue becomes larger than 1 and become unstable.

\begin{figure}[htb]
\begin{center}
\epsfysize=240pt\epsfbox{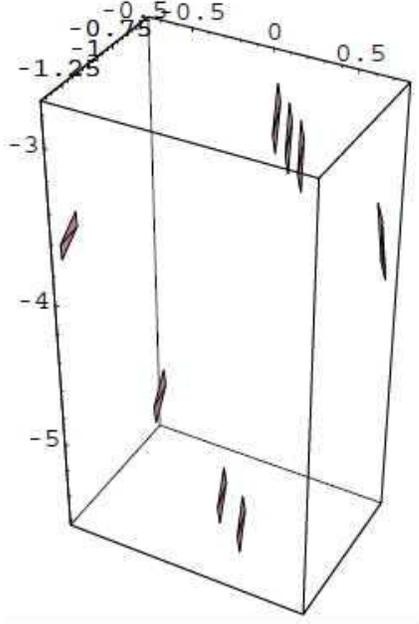}
\end{center}
\caption{Mapping of the tetrahedron spanned by 4 points $(x-0.1,y,z),(x,y-0.01,z),(x+0.1,y,z),(x,y+0.01,z)$ by the first derivative of the Poincar\'e mapping for $\tau=1546$,1553 to 1557, 1560 and 1564 from right to left. A shift of $\Delta z=-2$ occurs between $\tau=1555$ and 1556. }\label{polygon_all}
\end{figure}

In order to check whether the transition at $\tau=1554$ is smooth, we calculate the tetrahedron $\Delta \bf f$ produced by mapping of 4 points $(x-0.1,y,z),(x,y-0.01,z),(x+0.1,y,z),(x,y+0.01,z)$ by the Jacobi matrix  $\displaystyle \frac{\partial \bf f}{\partial{\bf x}}$  at $\tau=1546$, 1553 to 1557, 1560 and 1564. ( Fig.\ref{polygon_all} )  The area projected on the $x-y$ plane is nearly parallelogram . One of the eigenvalues of the first derivative of the Poincar\'e mapping is consistent with 0 and the rest are a pair of complex value in $1546\leq \tau\leq 1553$ and at $\tau=1554$ all the eigenvalues become real.  In the $z$ direction, a shift of $\Delta z=-2$ occurs between $\tau=1555$ and 1556, but on the projected $x-y$ plane the movement is smooth, which suggests that the orbit in $W^u(P_+)$ is smoothly connected in $W^s(P_-)$.

\subsection{The impulse-2a type control}
The impulse-2 type control is done by applying impulse at $x=-1$ as well as $x=1$. In this case, eigenvalues of the first derivative of the Poincar\'e mapping are real near the point where the 2nd impulse is applied to the system. The real largest eigenvalue crosses $\mu=1$ from above near the time when the 2nd pulse is applied.  The slope of the negative eigenvalue becomes flat at the same time.
The 1-d bifurcation diagram shown in Fig.\ref{impulse2_w}, we find 4 windows, but we present results of 2 windows.

\begin{figure}[htb]
\begin{center}
\epsfysize=160pt\epsfbox{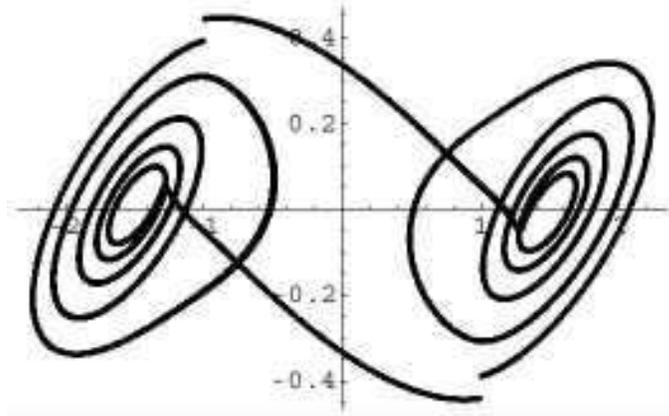}
\end{center}
\caption{The trajectory of the Chua's circuit impulse-2a type.}\label{impulse_2}\end{figure}

In the impulse 2a type control, we choose $\delta y=0.05$, whose
 controlled trajectory (the last 5000 steps of 50000 steps) is shown in Fig. \ref{impulse_2}.

\begin{figure}[htb]
\begin{minipage}[b]{0.47\linewidth}
\begin{center}
\epsfysize=120pt\epsfbox{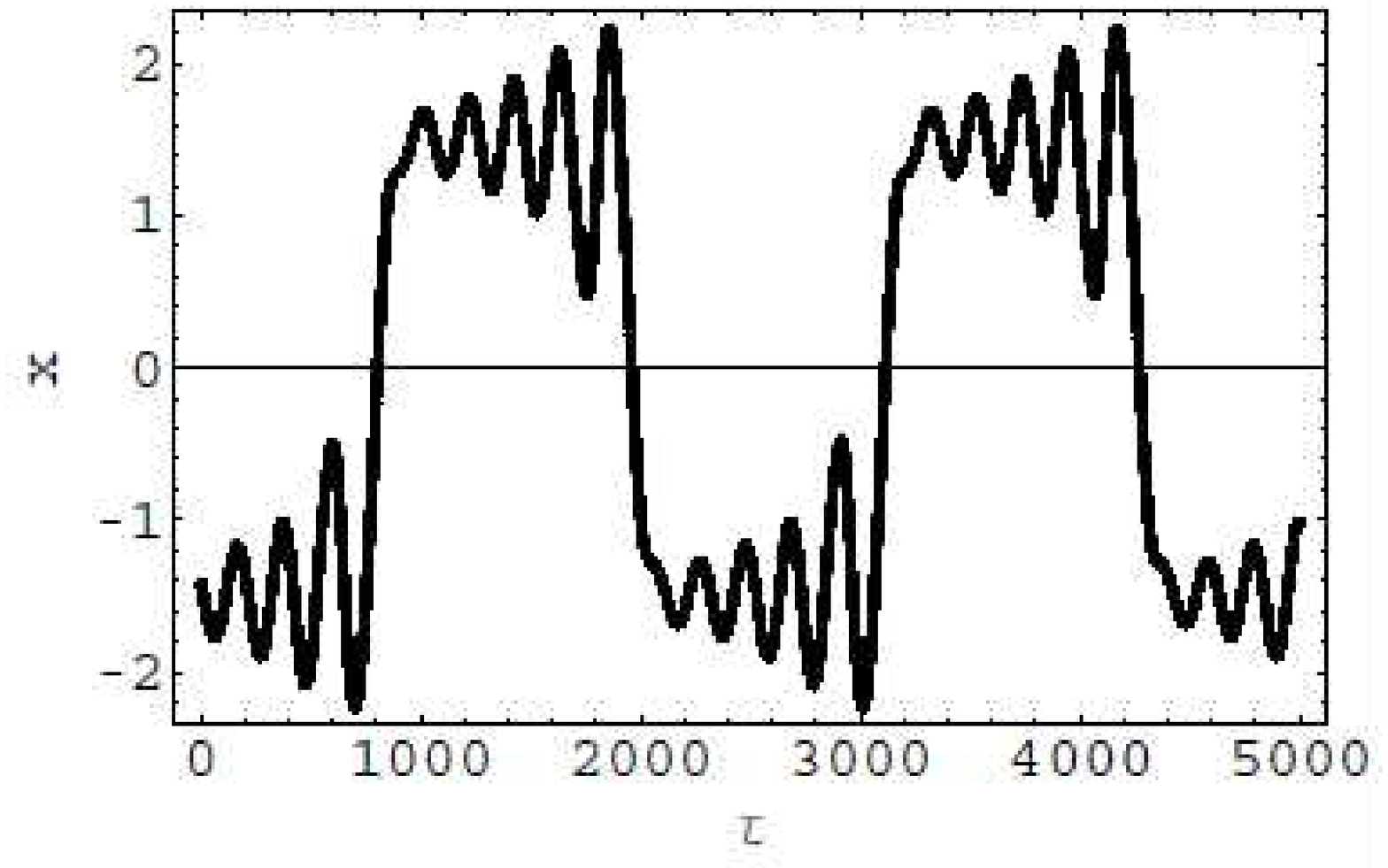}
\end{center}
\caption{The $x$ coordinate of the trajectory of impulse-2a type.}\label{impulse_2xn}
\end{minipage}
\hfill
\begin{minipage}[b]{0.47\linewidth}
\begin{center}
\epsfysize=120pt\epsfbox{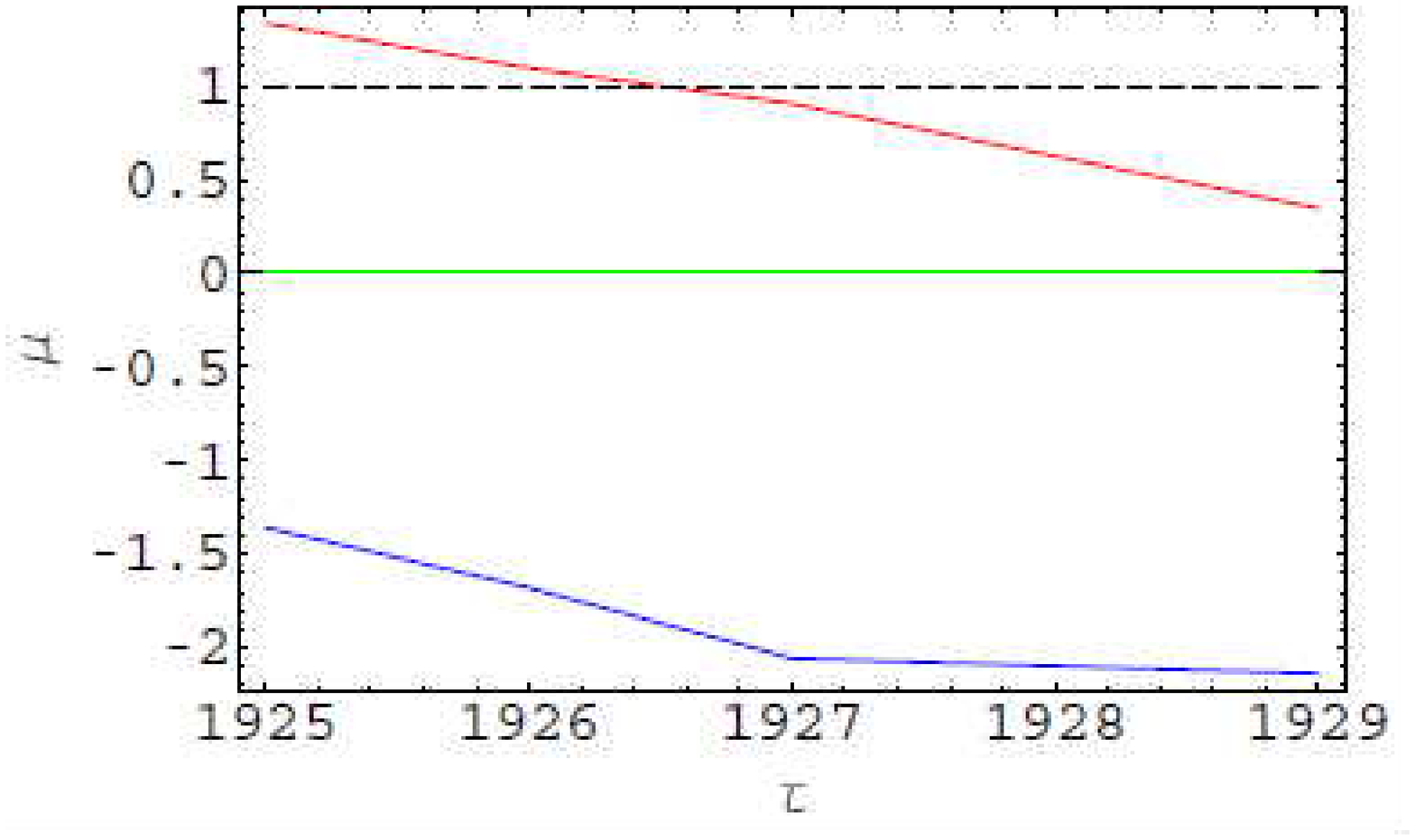}
\end{center}
\caption{Eigenvalues of the first derivative of the Poincar\'e mapping of the impulse-2a type.}\label{niiya2_mu}
\end{minipage}
\end{figure}

\begin{figure}[htb]
\begin{center}
\epsfysize=160pt\epsfbox{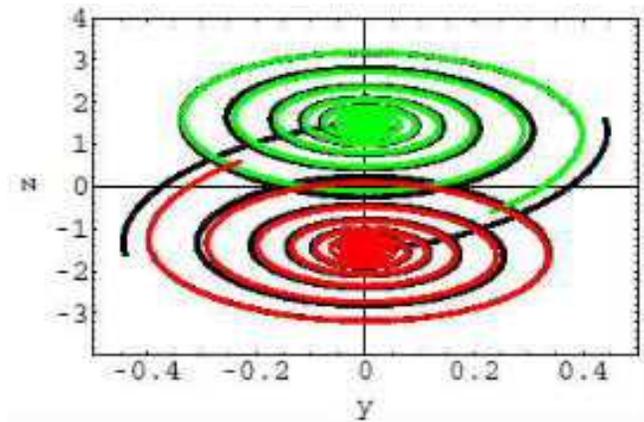}
\end{center}
\caption{Tangency of the controlled orbit (impulse-2a) and the two homoclinic orbits.}\label{newhouse2}
\end{figure}

The first derivative of the Poincar\'e mapping is calculated from $\tau=771$ when the $\delta y$ is applied to $\tau=1927$ when the subsequent  $\delta y$ is applied.  
In this region one of the three eigenvalues is consistent with 0 and the product of the rest two eigenvalues before $\tau=1927$ is smaller than -1.5 but after $\tau=1927$ it rapidly increases and becomes about 3 at $\tau=1937$.
The tangency of the unstable manifold $W^u(\Lambda_-)$ and the stable manifold $W^s(\Lambda_+)$ can be visualized by comparing the controlled trajectory in the $y-z$ plane and the homoclinic orbits started from the vicinity of the fixed point $P_+'=(-1.5+0.001,0,1.5)$ and $P_-'=(1.5-0.001,0,-1.5)$, which are shown in Fig.\ref{newhouse2}

\subsection{The impulse-2b type control}

\begin{figure}[htb]
\begin{center}
\epsfysize=120pt\epsfbox{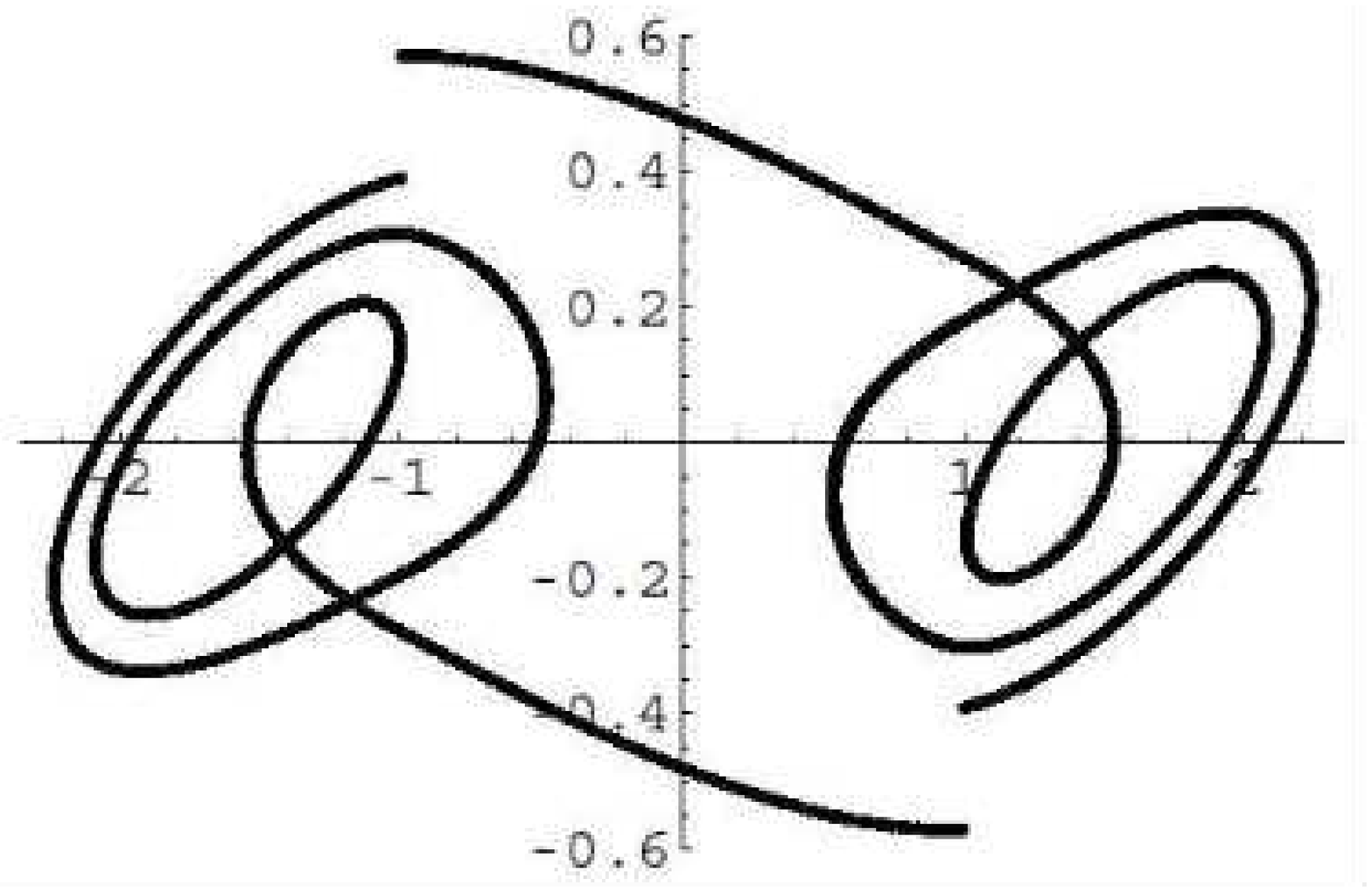}
\end{center}
\caption{The trajectory of Chua's circuit impulse-2b type.}\label{impulse2b}
\end{figure}

\begin{figure}[htb]
\begin{minipage}[b]{0.47\linewidth}
\begin{center}
\epsfysize=120pt\epsfbox{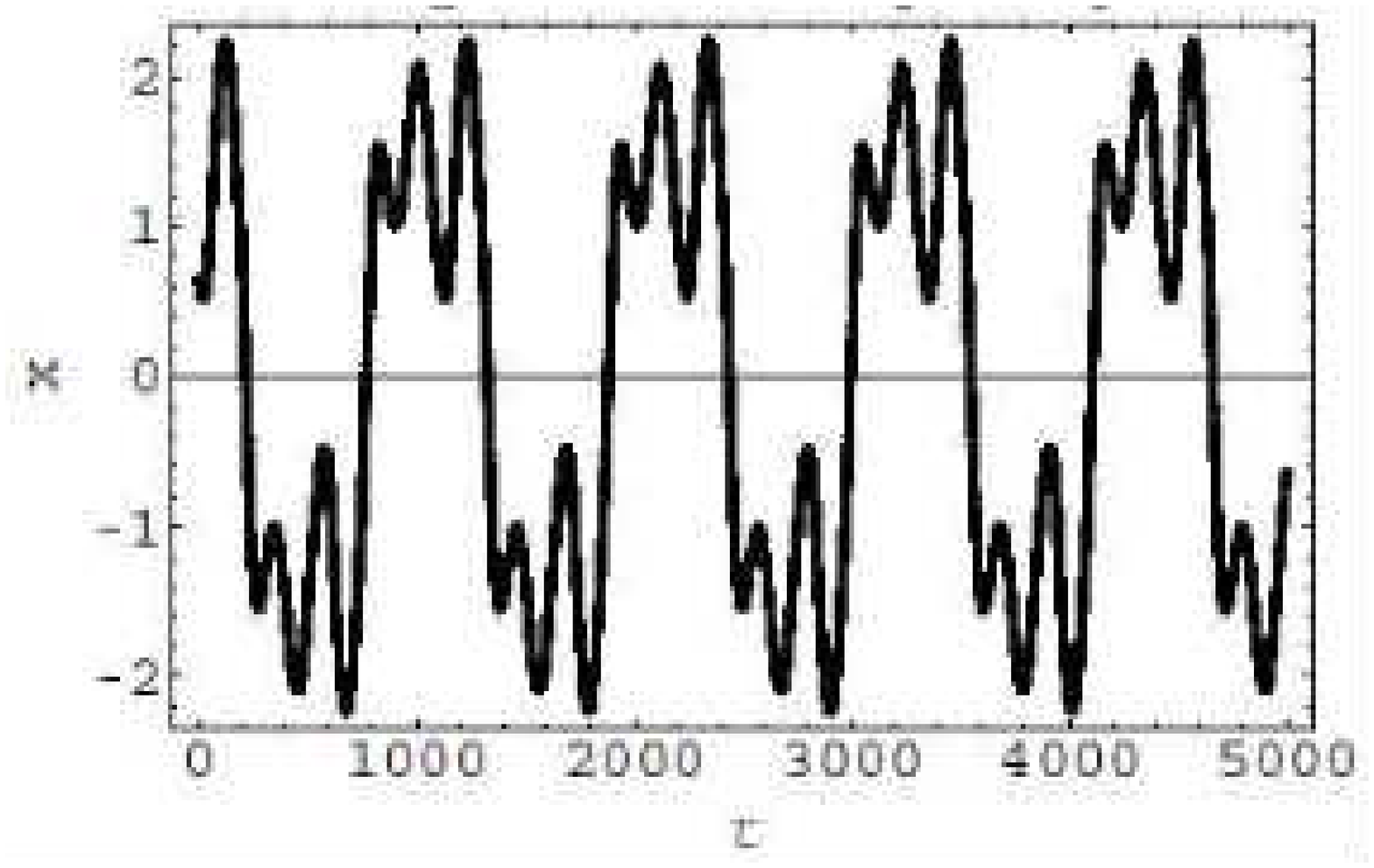}
\end{center}
\caption{The $x$ coordinate of the trajectory of impulse-2b type.}\label{impulse_2x}
\end{minipage}
\hfill
\begin{minipage}[b]{0.47\linewidth}
\begin{center}
\epsfysize=120pt\epsfbox{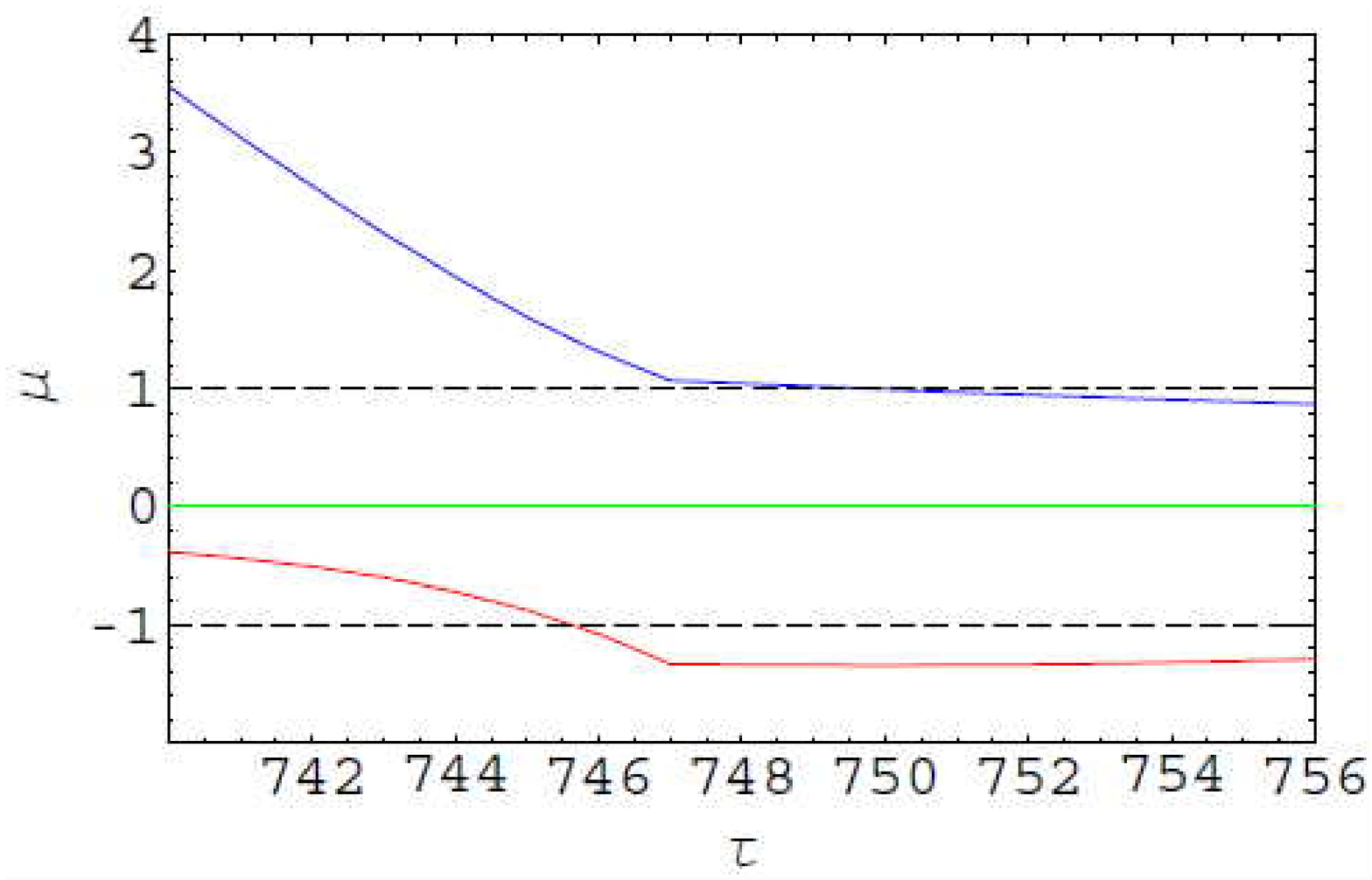}
\end{center}
\caption{Eigenvalues of the first derivative of the Poincar\'e mapping of the impulse-2b type.}\label{impulse_7mu}
\end{minipage}
\end{figure}

In the impulse-2b type we choose $\delta y=0.18$. The trajectory is shown in Fig.\ref{impulse2b} and the projection on the $x$-axis is shown in Fig.\ref{impulse_2x}.


The first derivative of the Poincar\'e mapping is calculated from $\tau=191$ when the $\delta y$ is applied to $\tau=748$ when the subsequent  $\delta y$ is applied.   Its eigenvalues are shown in Fig.\ref{impulse_7mu}. Product of two real eigenvalues are negative and smaller than -1.3 before $\tau=748$ but it rapidly approaches to -1 after $\tau=748$.
Products of the three eigenvalues after $\tau=748$ is constant ($0.38(1)\times 10^{-6}$), which means that the orbit is absorbed in the stable manifold $W^s(P_+)$.

\section{Drive control of the double scroll}

As in the forced oscillator, the chaotic system can be controlled by applying sinusoidal perturbation $F\sin\omega \tau$. In this case in the next feedback control, we choose $G=0.65$ so that the double scroll is produced.

\begin{equation}\label{chua_drive}
\left\{\begin{array}{l}
\frac{d}{d\tau}x=\alpha(y-x-g(x,m_0,m_1))\\
\frac{d}{d\tau}y=x-y+z\\
\frac{d}{d\tau}z=-\beta y+F\sin \omega\tau
\end{array}\right.
\end{equation}
We choose the angular frequency $\omega=\pi$ and define  the strength by looking at the 1-d bifurcation diagram as a function of $F$. Corresponding to the windows in the diagram of Fig.\ref{drive_w}, we consider following three types of drive control.

\begin{itemize}
\item Drive control-a : $F=1.4$

\item Drive control-b : $F=1.17$

\item Drive control-c : $F=1.7$ 
\end{itemize}

\subsection{The drive control-a}

\begin{figure}[htb]
\begin{minipage}[b]{0.47\linewidth}
\begin{center}
\epsfysize=120pt\epsfbox{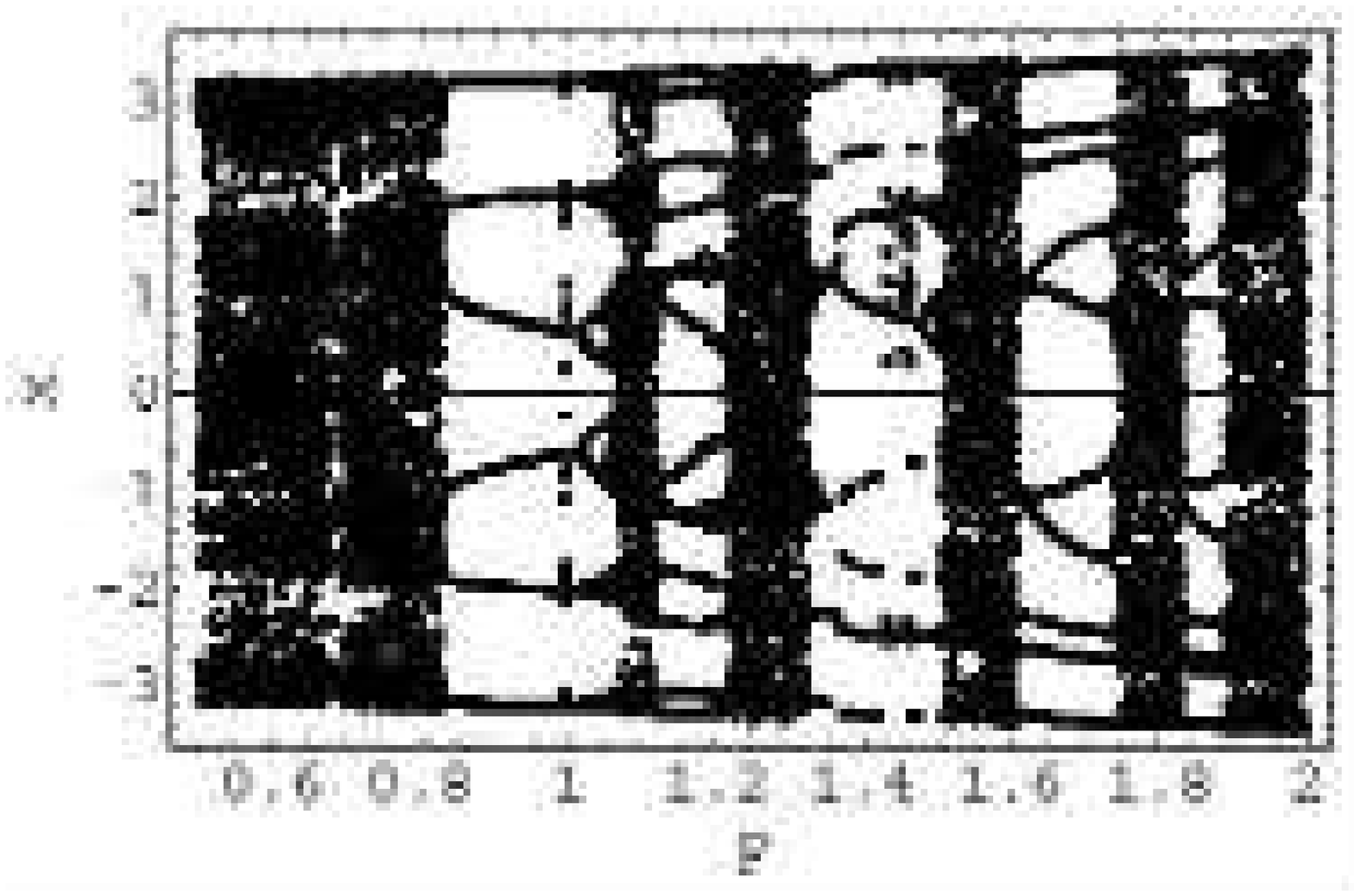}
\end{center}
\caption{The 1-d bifurcation diagram as a function of the amplitude $F$ of the drive control. }\label{drive_w}
\end{minipage}
\hfill
\begin{minipage}[b]{0.47\linewidth}
\begin{center}
\epsfysize=120pt\epsfbox{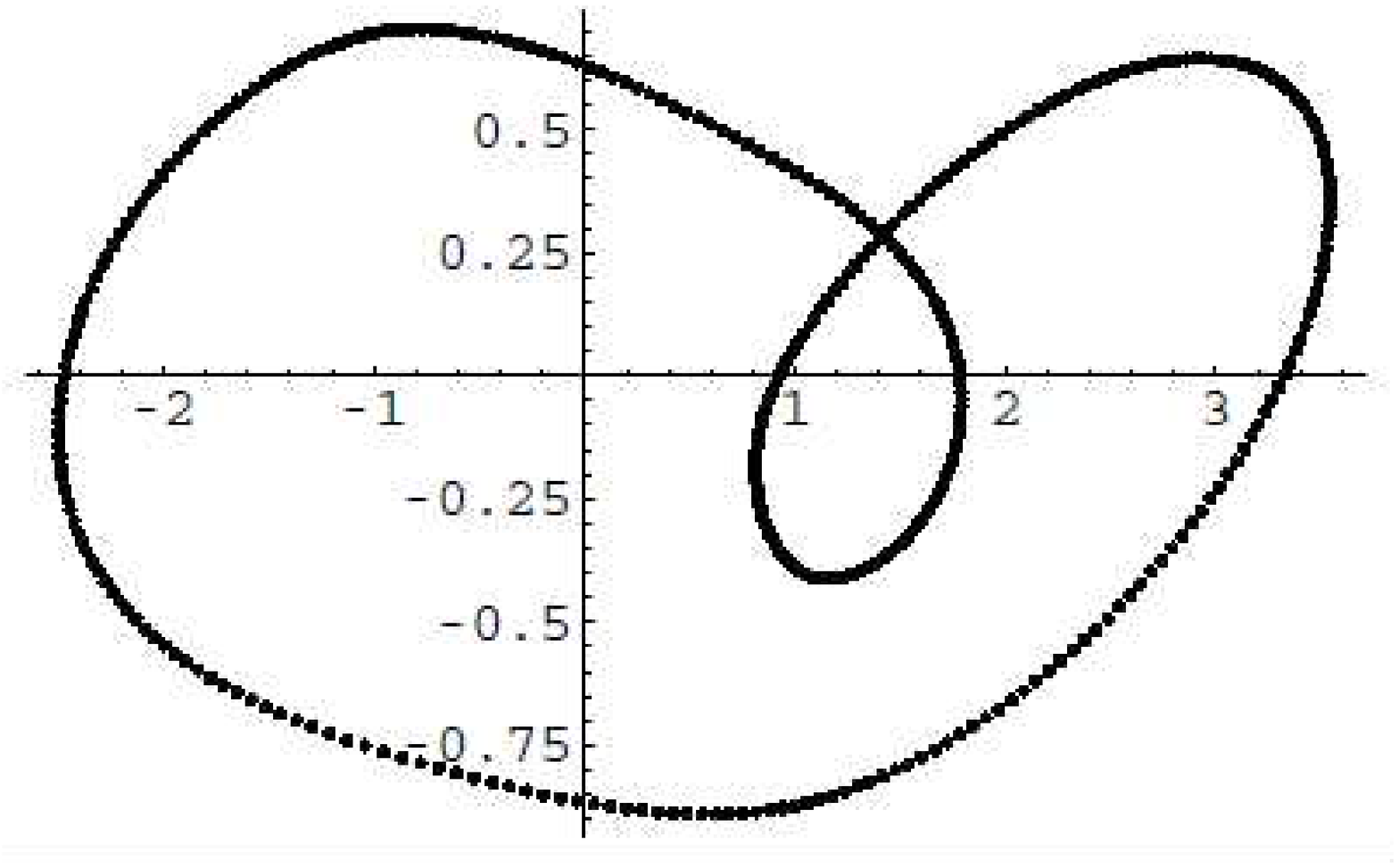}
\end{center}
\caption{The drive control-a of the double scroll.}\label{drive_1p4}
\end{minipage}
\end{figure}

\begin{figure}[htb]
\begin{center}
\epsfysize=120pt\epsfbox{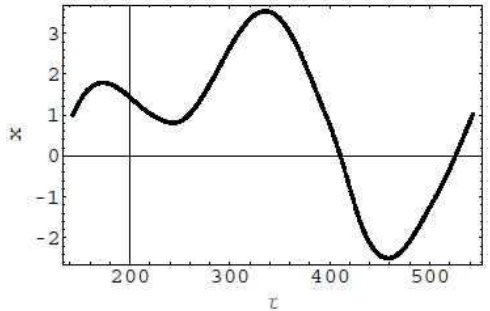}
\end{center}
\caption{The $x$ coordinate of the trajectory of drive control-a type.}\label{drive_x}
\end{figure}
A simulation data of $F=1.4, \omega=\pi$ is shown in Fig.\ref{drive_1p4}.
The first derivative of the Poincar\'e mapping is calculated from $\tau=143$ to $\tau=543$ where the initial point is revisited. 
Its eigenvalues change from a pair of complex and a real to three real periodically. Similar behavior was observed\cite{fur03} in the control of the Lorenz equation\cite{Lor63}.
\begin{figure}[htb]
\begin{minipage}[b]{0.47\linewidth}
\begin{center}
\epsfysize=120pt\epsfbox{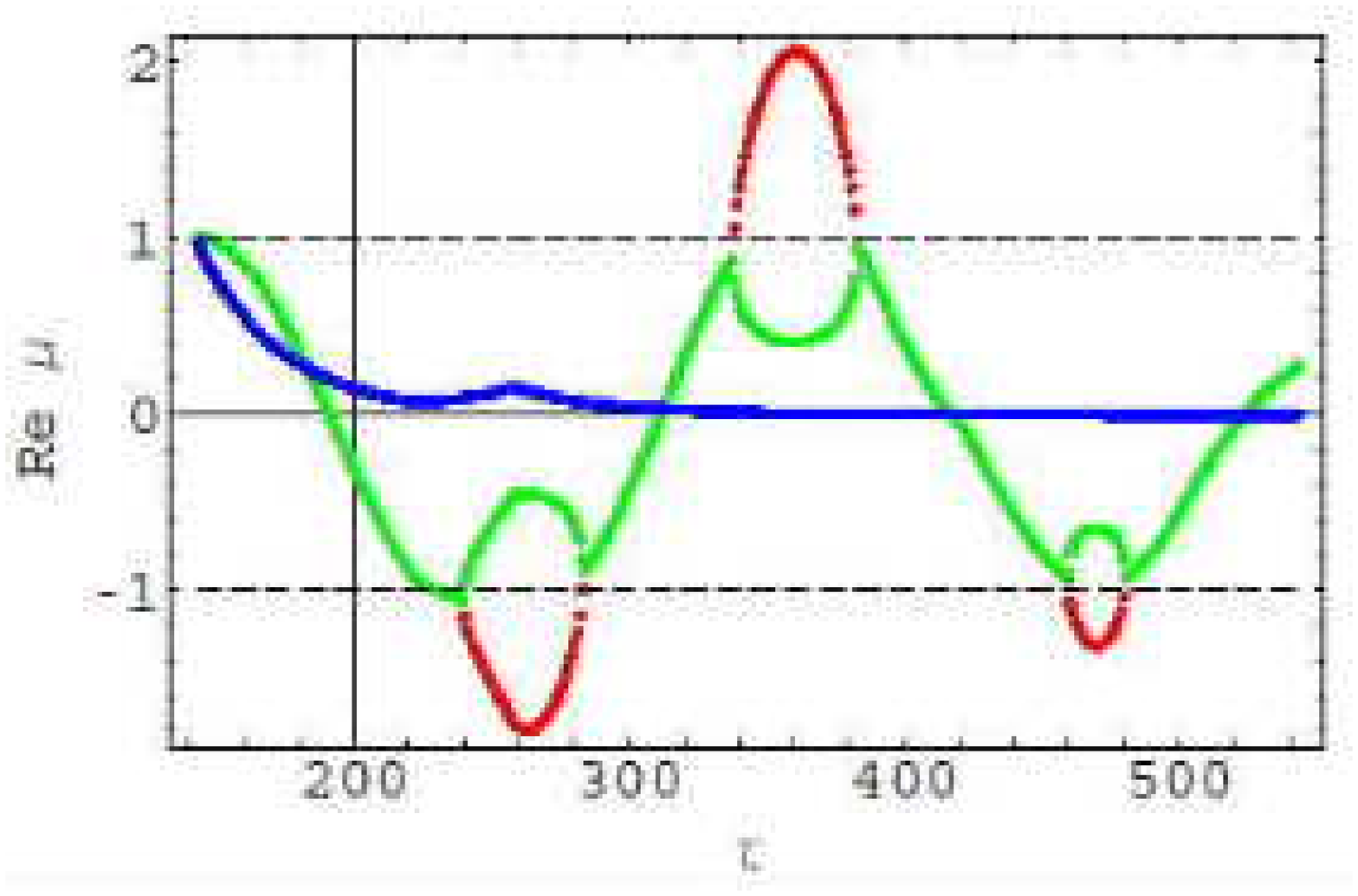}
\end{center}
\caption{The real part of eigenvalues drive control-a type.}\label{drive_re_mu}
\end{minipage}
\hfill
\begin{minipage}[b]{0.47\linewidth}
\begin{center}
\epsfysize=120pt\epsfbox{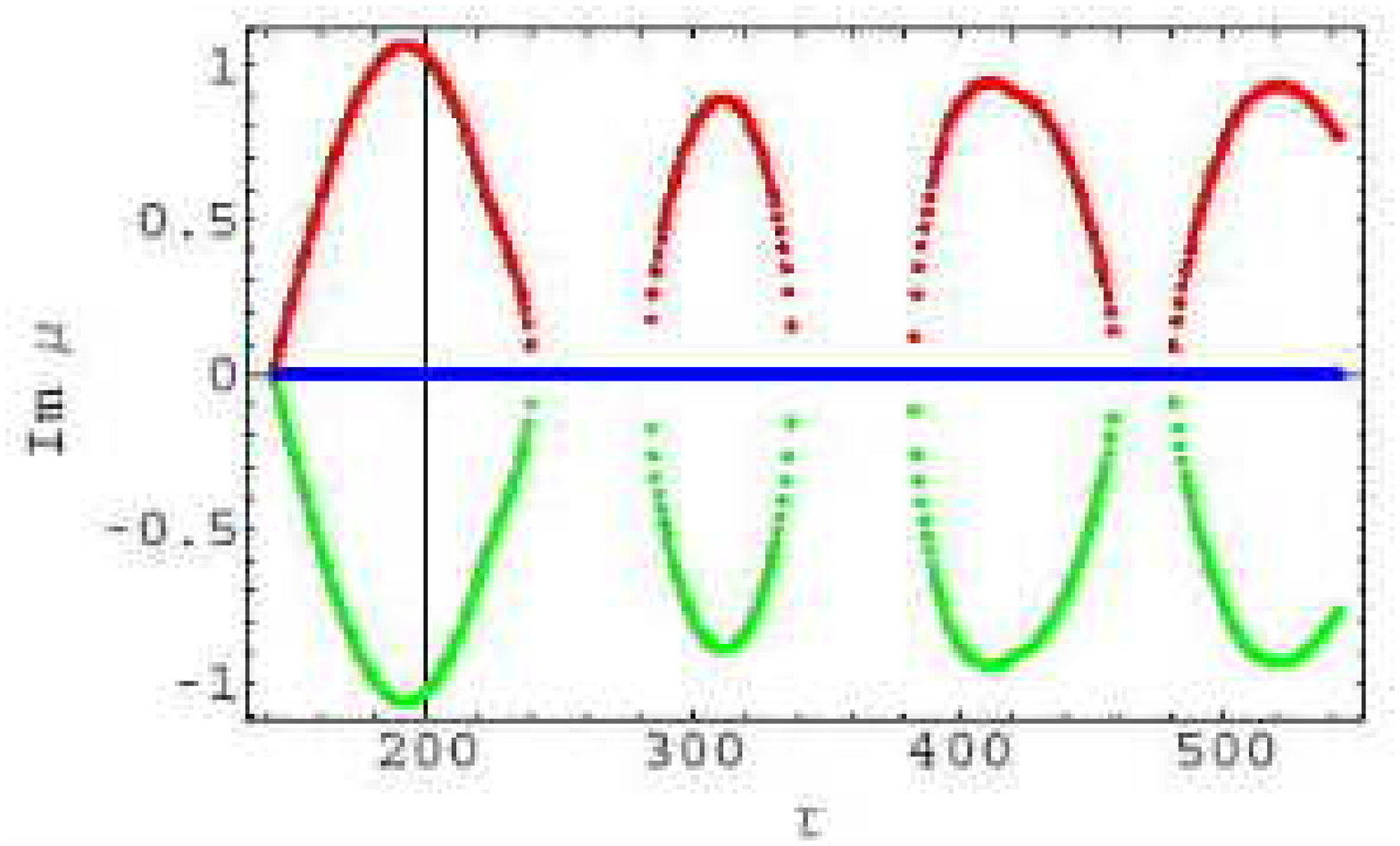}
\end{center}
\caption{The imaginary part of eigenvalues drive control-a type.}\label{drive_im_mu}
\end{minipage}
\end{figure}

In the interval from $\tau=338$ to 382, the diffeomorphism is essentially in 2-dimensional plane and the two real eigenvalues satisfy the condition of the Newhouse region $\displaystyle \rho<1<\lambda<\frac{1}{\rho}$ and the orbit is absorped into the stable periodic orbit. In the period of $387<\tau<405$ products of the two eigenvalues $\rho\lambda$ are almost constant 0.91(1).

\subsection{The drive control-b}
\begin{figure}[htb]
\begin{minipage}[b]{0.47\linewidth}
\begin{center}
\epsfysize=120pt\epsfbox{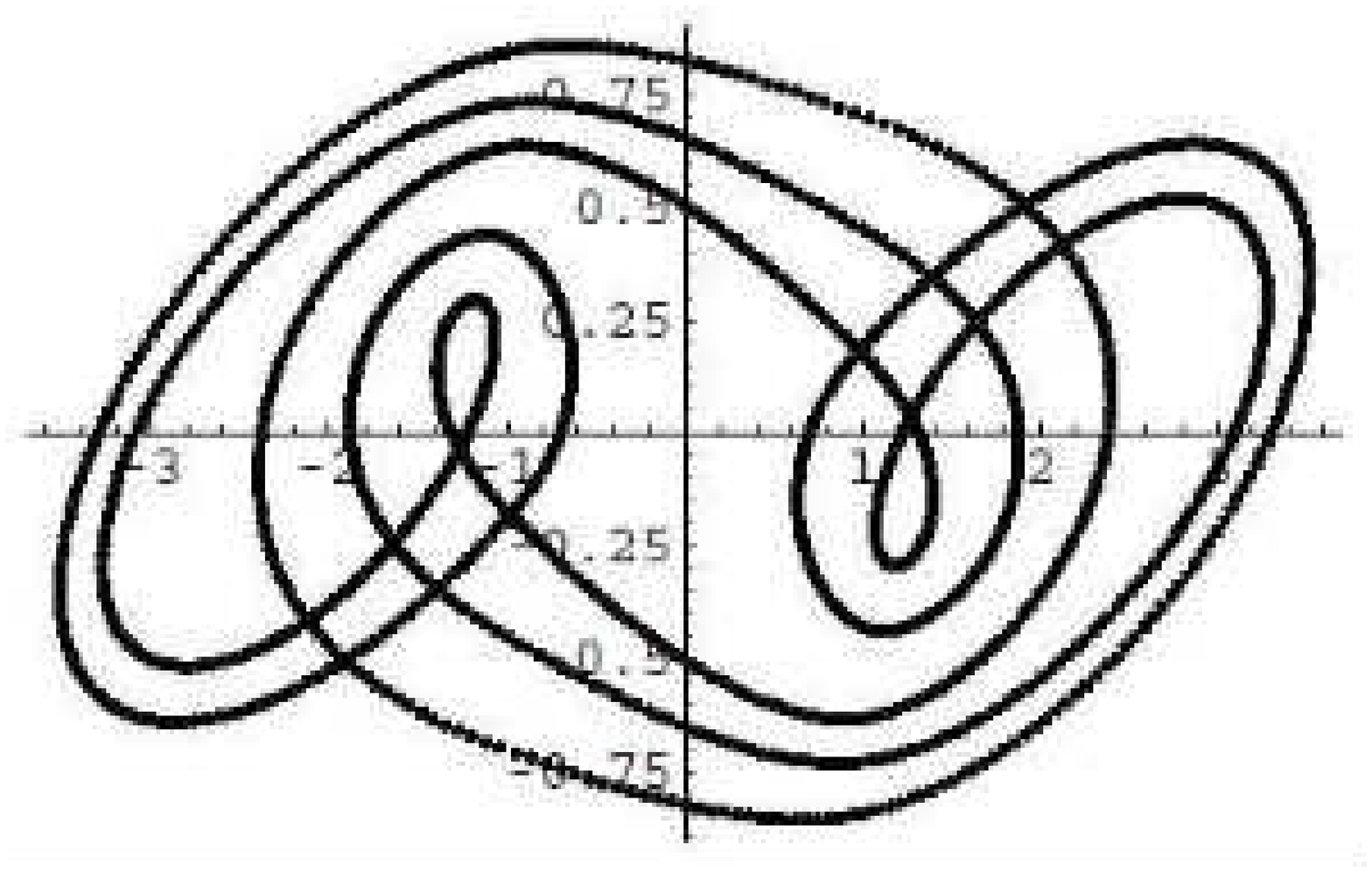}
\end{center}
\caption{The drive control-b of the double scroll.}\label{drive_1p17}
\end{minipage}
\hfill
\begin{minipage}[b]{0.47\linewidth}
\begin{center}
\epsfysize=120pt\epsfbox{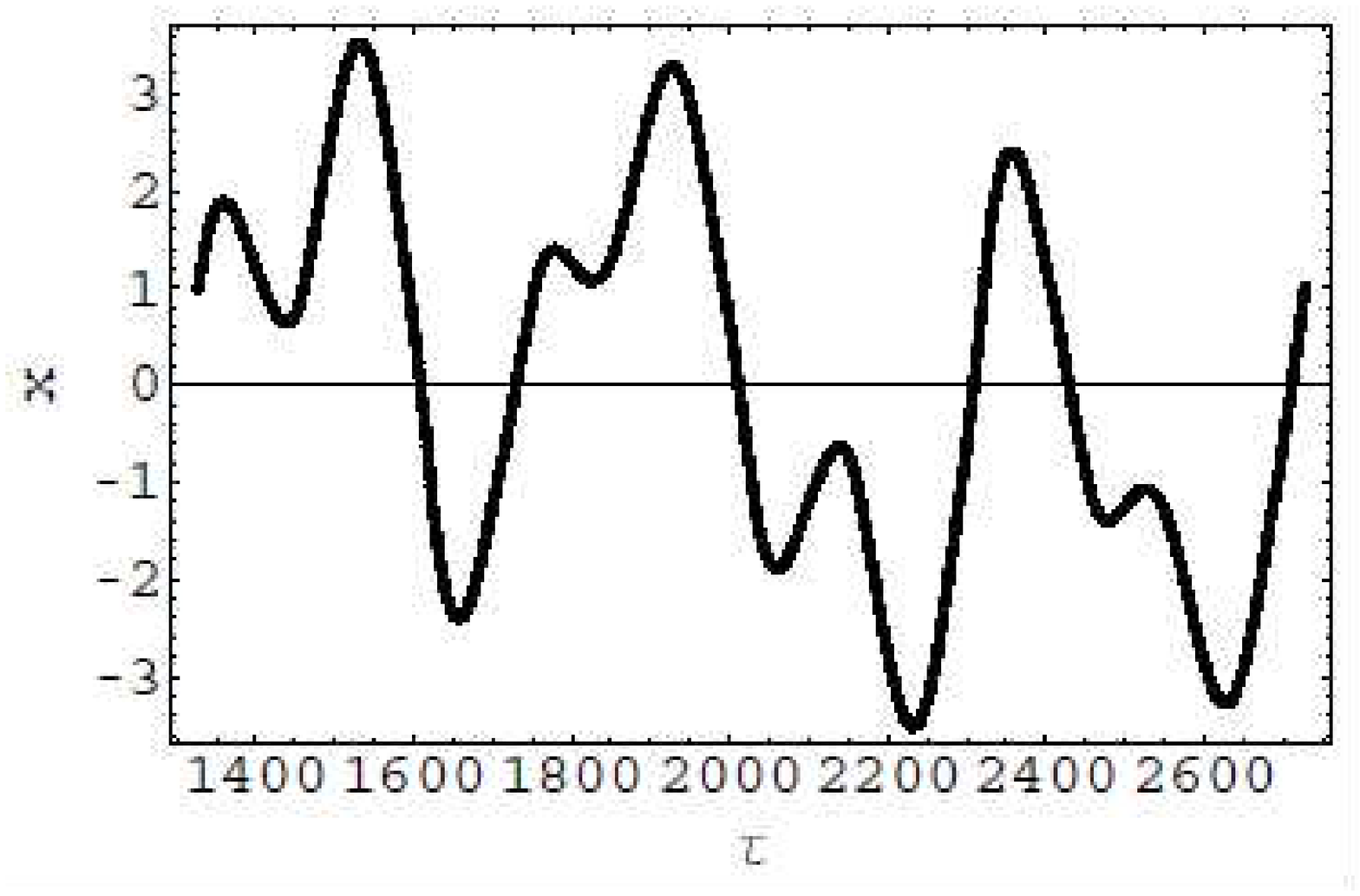}
\end{center}
\caption{The $x$ coordinate of the trajectory of drive control-b type.}\label{driveb_x}
\end{minipage}
\end{figure}

A simulation data of $F=1.17, \omega=\pi$ is shown in Fig.\ref{drive_1p17}. 

 The first derivative of the Poincar\'e mapping is calculated in the period from $\tau=1328$ to $\tau=2728$. 
Its eigenvalues change from a pair of complex and a real to three real periodically. 
The third eigenvalue is small and although the product of the complex eigenvalues become larger than 1, the absolute value of the product of three eigenvalues   is small and the periodic trajectory is realized. Difference from the drive control-a is that the contraction mapping here is 3-dimensional.

\subsection{The drive control-c}

\begin{figure}[htb]
\begin{minipage}[b]{0.47\linewidth}
\begin{center}
\epsfysize=120pt\epsfbox{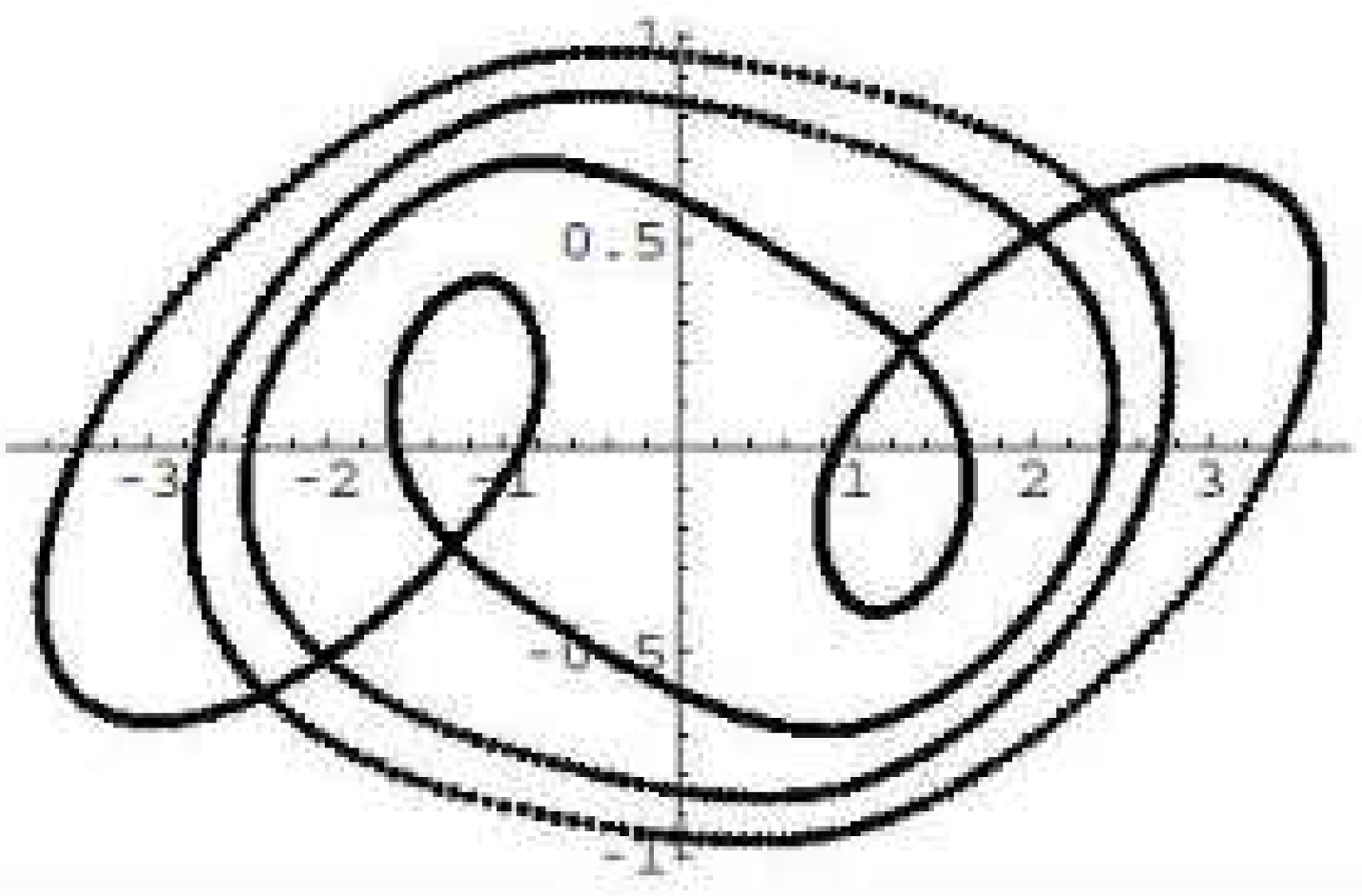}
\end{center}
\caption{The drive control-c of the double scroll.}\label{drive_1p7}
\end{minipage}
\hfill
\begin{minipage}[b]{0.47\linewidth}
\begin{center}
\epsfysize=120pt\epsfbox{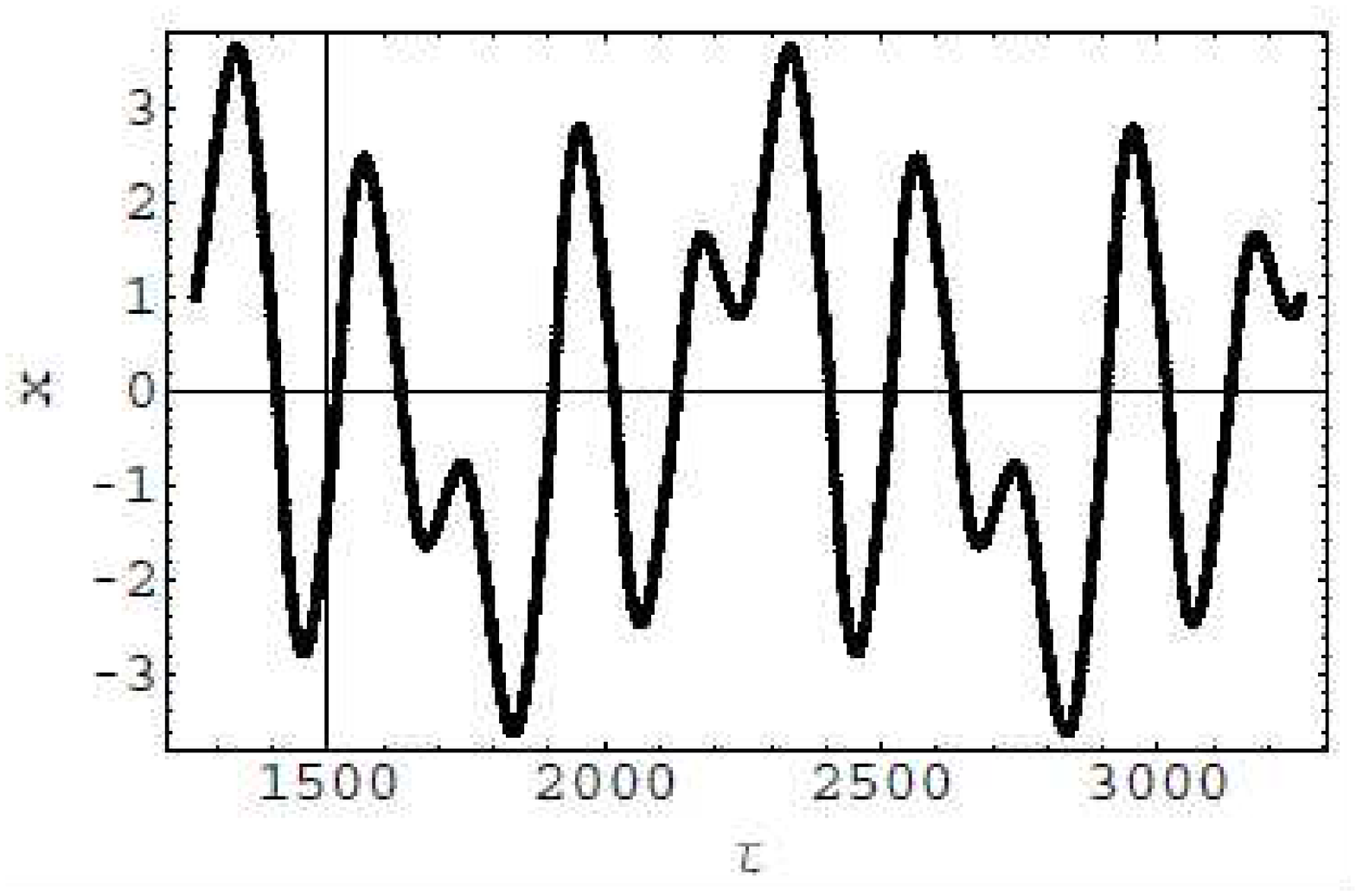}
\end{center}
\caption{The $x$ coordinate of the trajectory of drive control-c type.}\label{drivec_x}
\end{minipage}
\end{figure}

A simulation data of $F=1.7, \omega=\pi$ is shown in Fig.\ref{drive_1p7}. 
The first derivative of the Poincar\'e mapping is calculated in the period between $\tau=1259$ to $\tau=3259$. 
One of its eigenvalues is consistent with zero. Interchanges of a pair of complex to real ones occur periodically. The absolute value of the product of real eigenvalues $\rho\lambda$ exceeds 1 in the middle of the period but as it changes to a complex pair it is smaller than 1. Stretching in one direction and contracting in another direction occur when the eigenvalues are real and folding occurs when the eigenvalues are complex. 
Since the amplitude of the driving force is large, the oscillatory pattern of the complex eigenvalues is denser than that of the drive control-b type.

\section{Feedback control of the double scroll}
The application of feedback control is relatively simple. 
We choose the feedback time delay to be 1$d\tau$ and simulate the following equation.
\begin{equation}\label{chua_feedbk}
\left\{\begin{array}{l}
\frac{d}{d\tau}x=\alpha(y-x-g(x,m_0,m_1))\\
\frac{d}{d\tau}y=x-y+z \\
\frac{d}{d\tau}z=-\beta y+c \beta y[\tau]
\end{array}\right.
\end{equation}

\begin{figure}[htb]
\begin{minipage}[b]{0.47\linewidth}
\begin{center}
\epsfysize=120pt\epsfbox{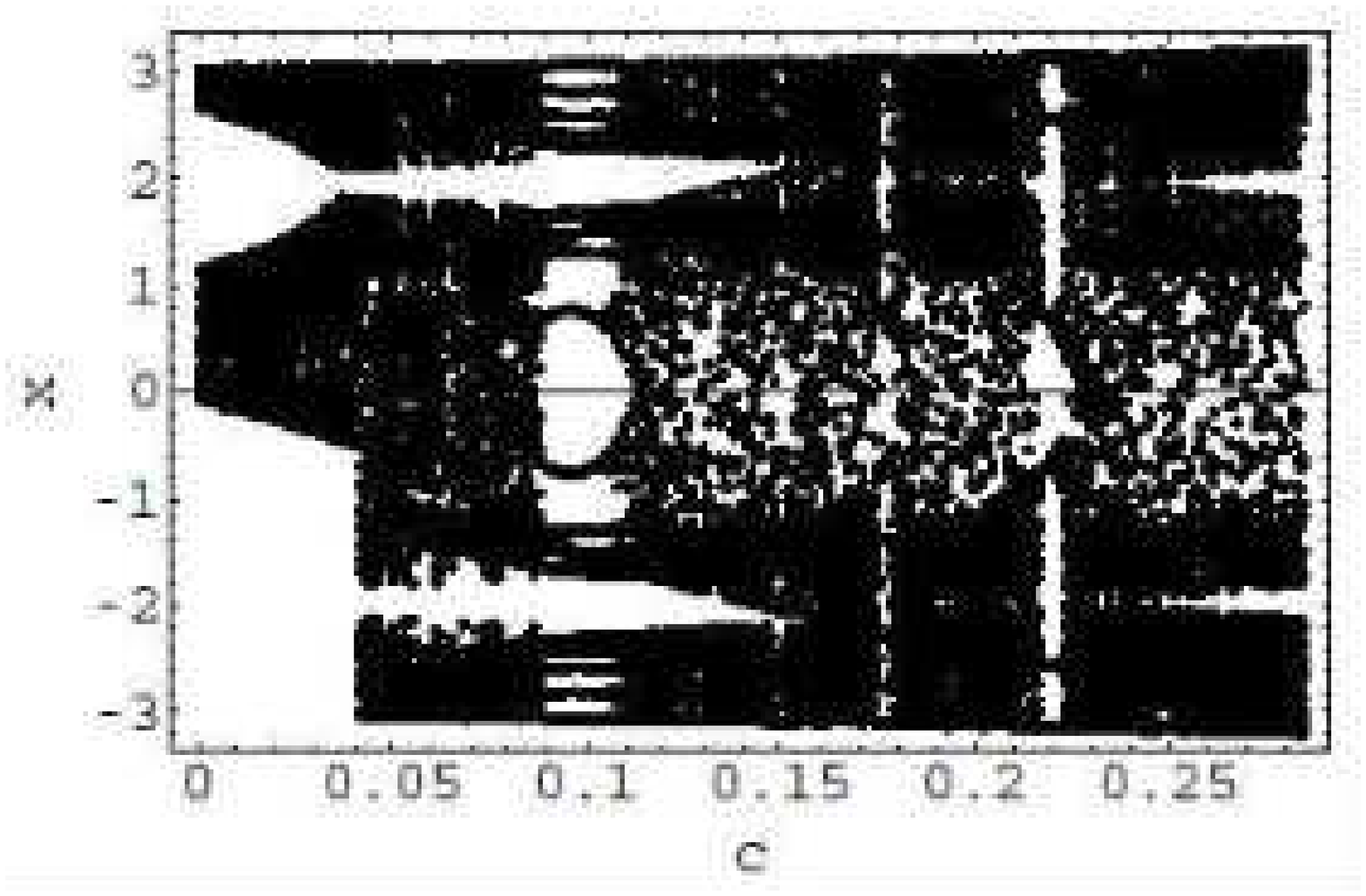}
\end{center}
\caption{The 1-d bifurcation diagram as a function of the parameter $c$ of the feedback control.}\label{feedbk_w}
\end{minipage}
\hfill
\begin{minipage}[b]{0.47\linewidth}
\begin{center}
\epsfysize=120pt\epsfbox{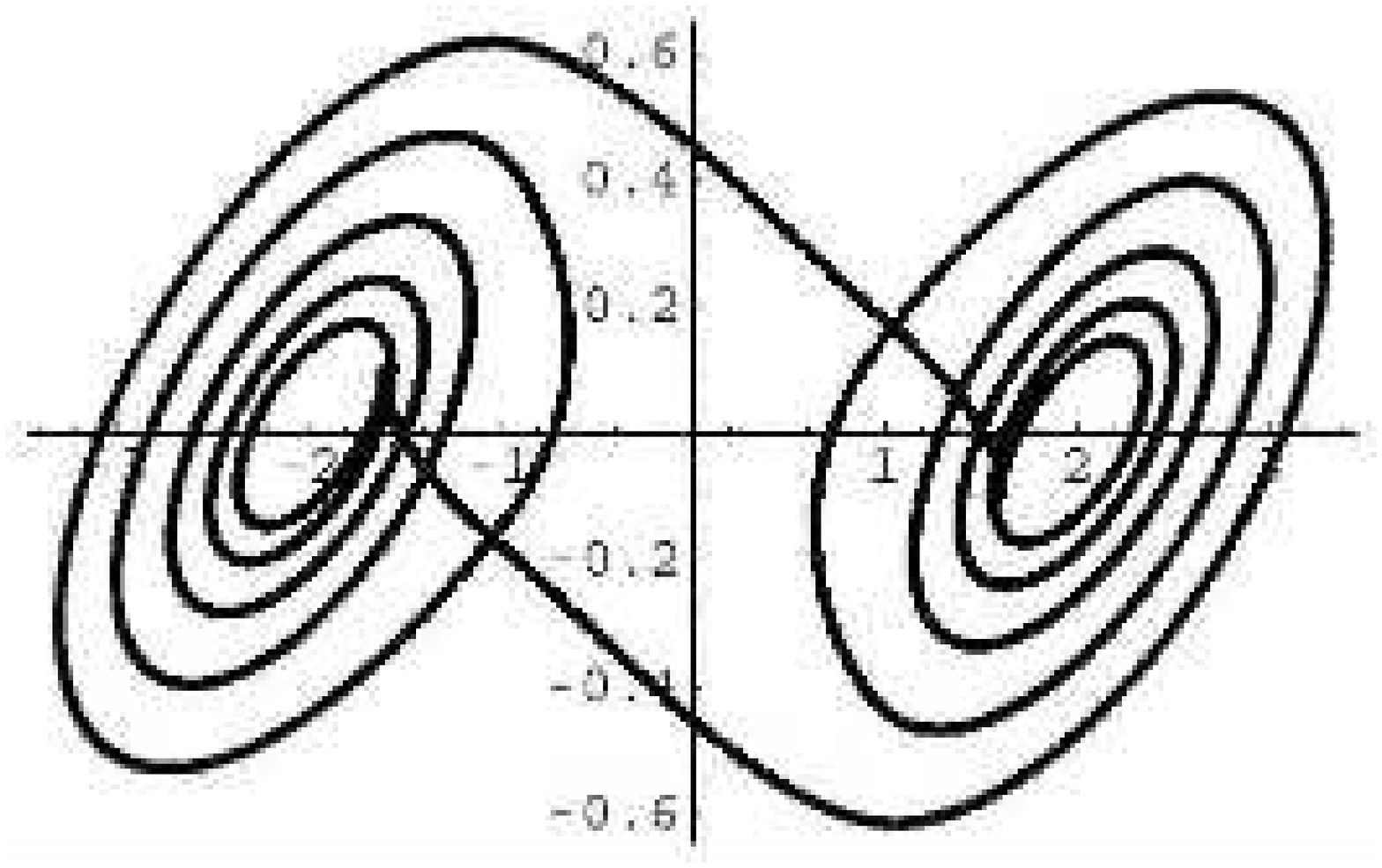}
\end{center}
\caption{The feedback control of the double scroll.}\label{feedback_p1}
\end{minipage}
\end{figure}

We select the strength $c$ by looking at the window in the 1-d bifurcation pattern as a function of $c$, as shown in Fig.\ref{feedbk_w}. 
The trajectory obtained by using $c=0.1$ is shown in Fig.\ref{feedback_p1}. This pattern is similar to the heteroclinic orbit shown in \cite{mc87}.

\begin{figure}[htb]
\begin{center}
\epsfysize=120pt\epsfbox{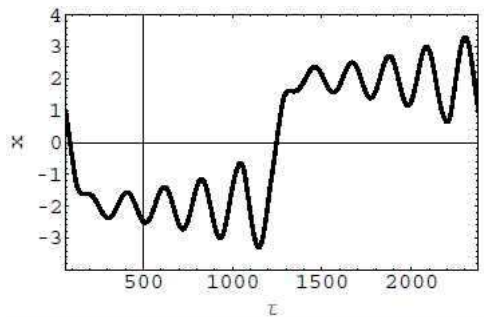}
\end{center}
\caption{The $x$ coordinate of the trajectory of feedback control type.}\label{feedbk_x}
\end{figure}

The first derivative of the Poincar\'e mapping is calculated from $\tau=64$ to $\tau=2676$ where the initial point is revisited. The $x$ coordinate of the trajectory is shown in Fig.\ref{feedbk_x}.
The eigenvalues of the first derivative of the Poincar\'e mapping changes from a pair of complex and a real to three real periodically. Their behavior is similar to that of the drive control-c type. 

\section{Impulse control of the spiral}

\begin{figure}[htb]
\begin{minipage}[b]{0.47\linewidth}
\begin{center}
\epsfysize=120pt\epsfbox{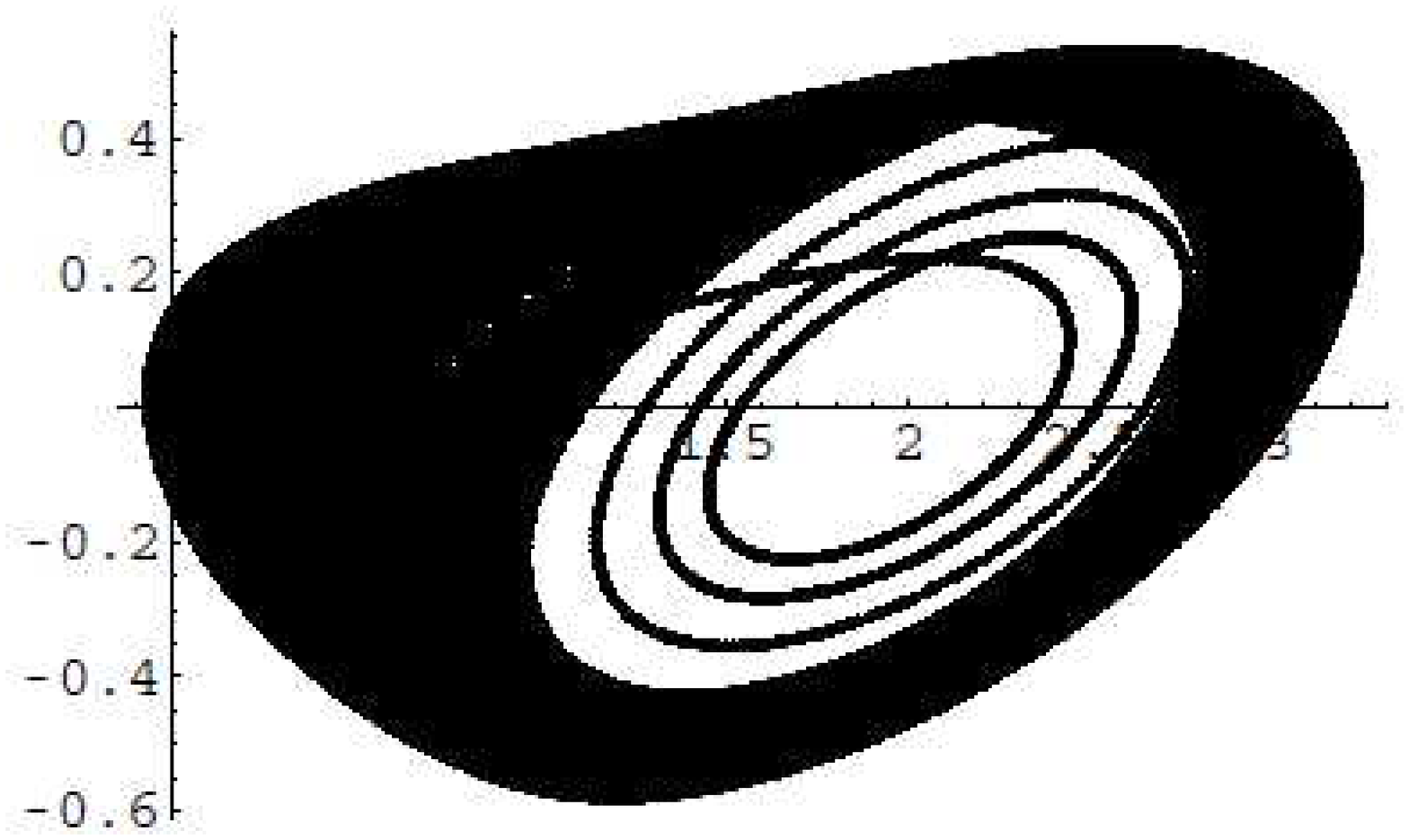}
\end{center}
\caption{The trajectory of the spiral without impulse control. }\label{spiral_chaos}
\end{minipage}
\hfill
\begin{minipage}[b]{0.47\linewidth}
\begin{center}
\epsfysize=120pt\epsfbox{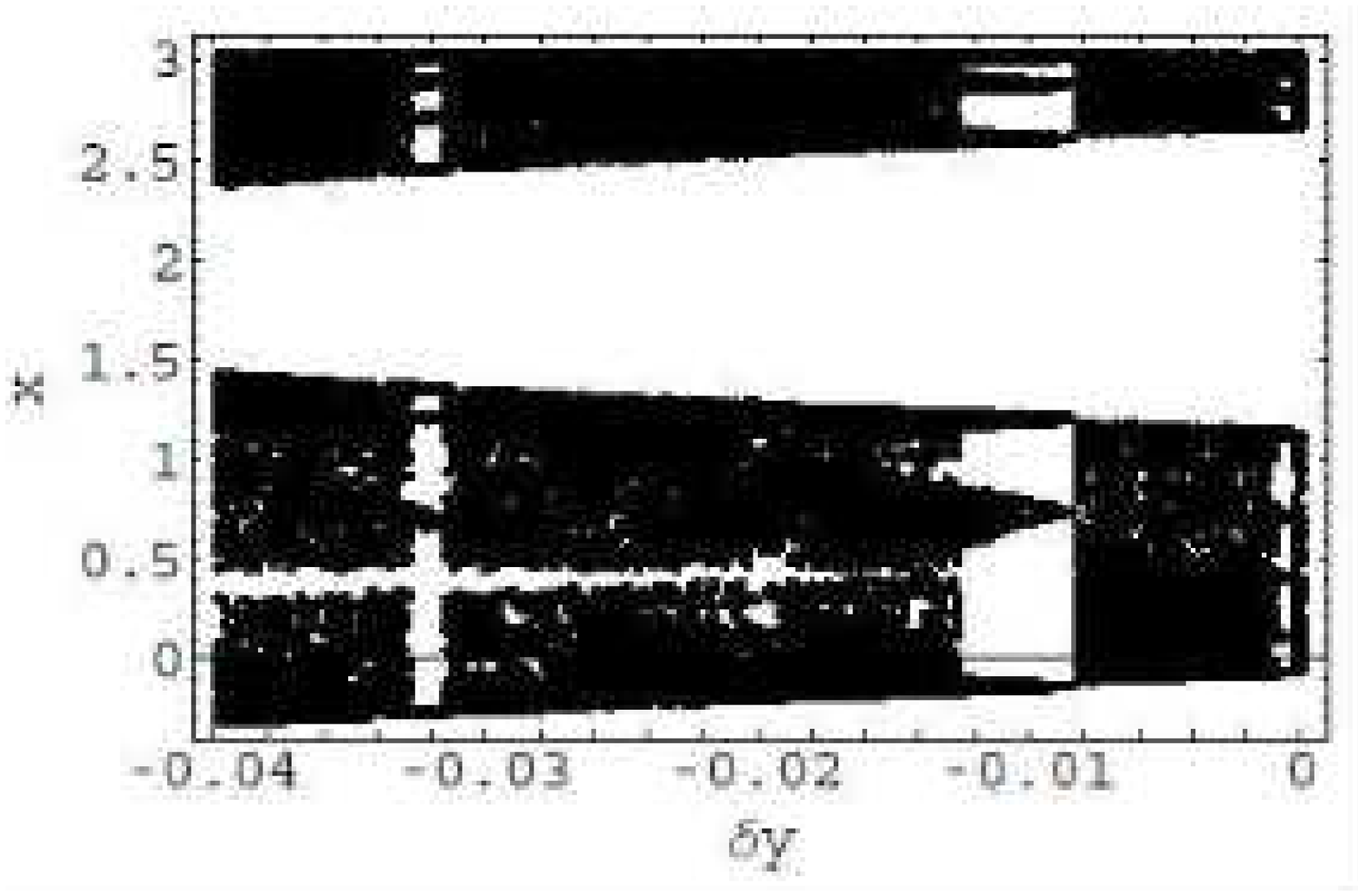}
\end{center}
\caption{The 1-d bifurcation diagram as the function of the height of the impulse.}\label{spiral_w}
\end{minipage}
\end{figure}

\begin{figure}[htb]
\begin{minipage}[b]{0.47\linewidth}
\begin{center}
\epsfysize=120pt\epsfbox{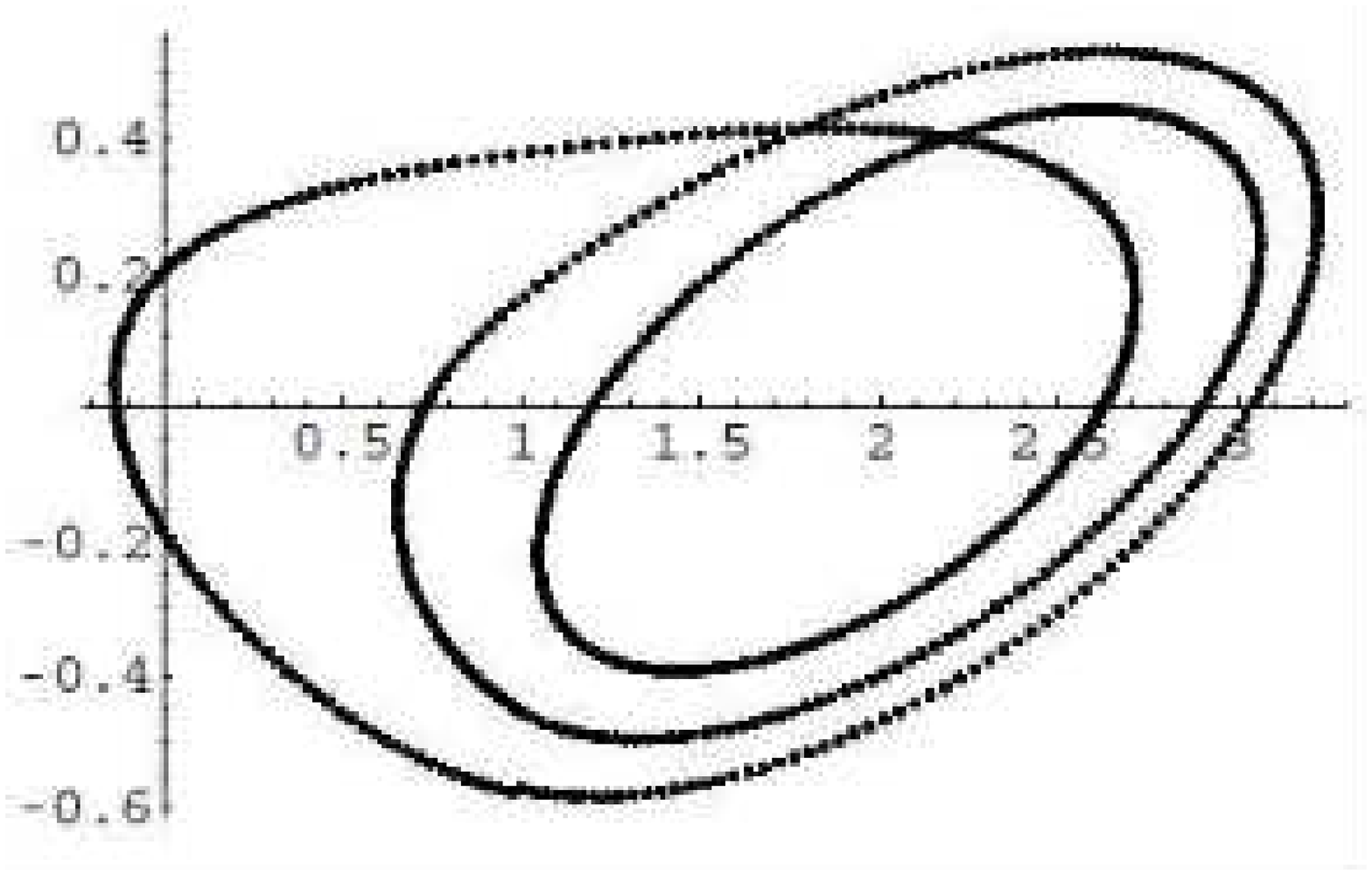}
\end{center}
\caption{The trajectory of spiral impulse-1 type. }\label{spiral_0088}
\end{minipage}
\hfill
\begin{minipage}[b]{0.47\linewidth}
\begin{center}
\epsfysize=120pt\epsfbox{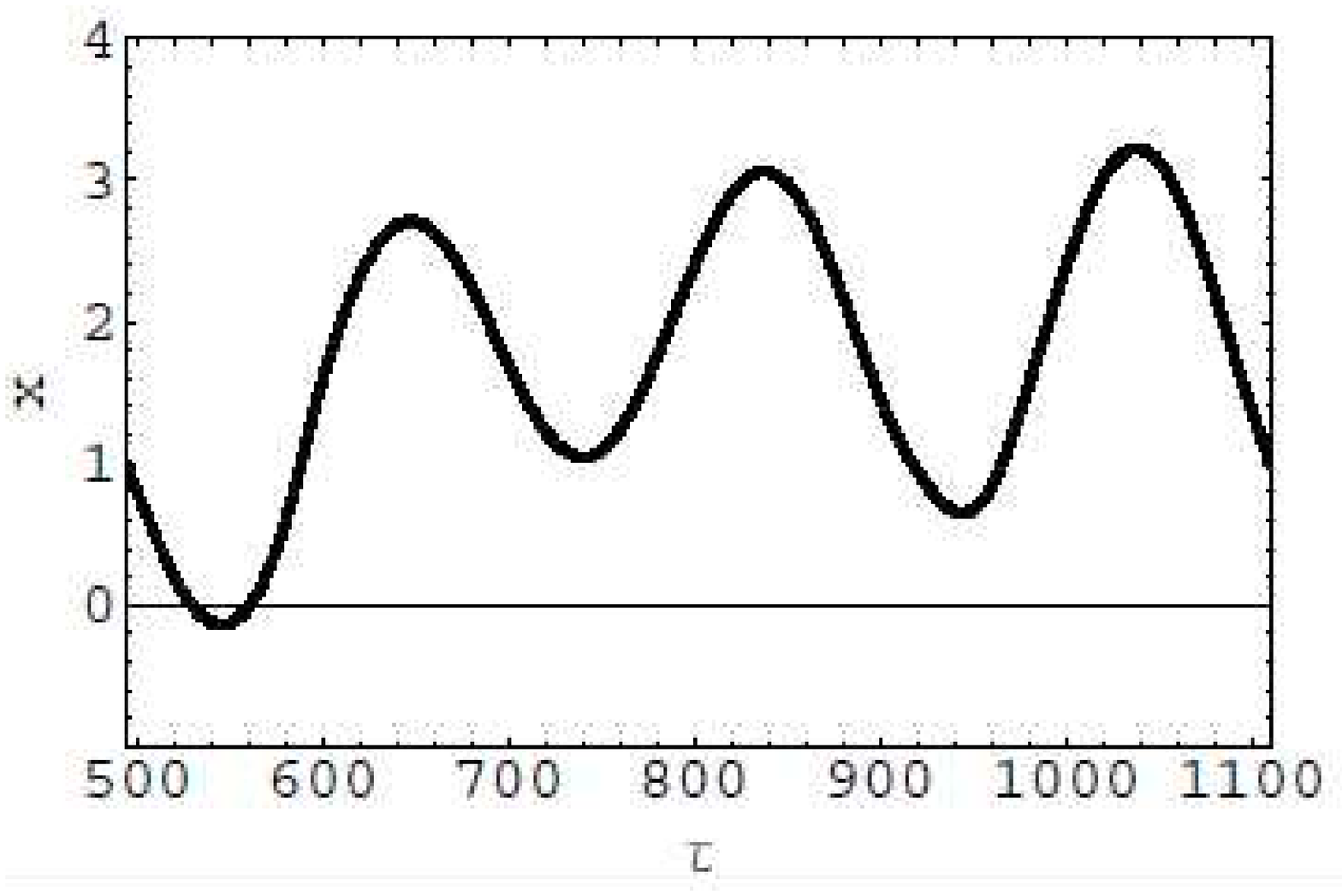}
\end{center}
\caption{The $x$ coordinate of the trajectory of the impulse-1 type.}\label{impulse_ex}
\end{minipage}
\end{figure}

The chaotic spiral can be obtained for the same parameters as those of double scroll written after eq.(\ref{chua_eq}) except $G=0.65$. The trajectory is shown in Fig.\ref{spiral_chaos}. We apply the impulse-1 type whose strength is fixed by looking for windows in the 1-d bifurcation diagram as a function of the pulse strength shown in Fig.\ref{spiral_w}. Observing that a crisis occurs at $\delta y\sim -0.008$, we apply the following control:
\begin{itemize}
\item Impulse-1 type: When $x$ passes 1 from $x_n>x_{n+1}$ and $y<-0.5$ shift $y$ to $y-0.0088$.
\end{itemize}
The Fig.\ref{spiral_chaos} shows the chaotic pattern and the Fig.\ref{spiral_0088} shows the result of impulse-1 type control. 

\begin{figure}[htb]
\begin{center}
\epsfysize=120pt\epsfbox{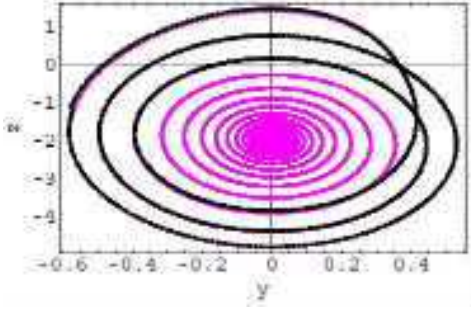}
\end{center}
\caption{Tangency of the impulse-1 type controlled orbit and the homoclinic orbit.}\label{spiral_hclnk}
\end{figure}

Although 2 pulses are triggered in the control of the double scroll, where heteroclinic tangency occurs, only one pulse is triggered in the impulse-1 type control, where homoclinic tangency occurs.  The crossing of the orbit on $W^u(\Lambda_-)$ and that on $W^s(\Lambda_-)$ occurs at the fixed point $P_-$ and the shift of the orbit on $W^u(\Lambda_-)$ into the region where countable periodic orbits exist makes the orbit ultimately tangential to $W^s(\Lambda_-)$ and the periodic orbit is realized.
The controlled trajectory in the $y-z$ plane and the homoclinic orbit started from the vicinity of the fixed point $P_-'=(2.0-0.00001,0,-2.0)$ are shown in Fig.\ref{spiral_hclnk}. 
The $x$ coordinate of the 3-cycle trajectory is shown in Fig.\ref{impulse_ex}. 

The eigenvalues depend on the initial condition. But the presence of the stretching, contraction and rotation (bending) is the same. The mapping of the tetrahedron $\Delta\bf f$ is almost parallel to the $x-z$ plane and the shift of $\Delta z$ observed in the double scroll is absent.

\section{Discussion and outlook}
We showed that the Shil'nikov's chaos in the Chua's circuit can be controlled by
1) imposing impulse or 2) applying phase synchronization or 3) performing feedback. The strength of the perturbation was defined by the position of windows in the 1-d bifurcation diagram.  The perturbation on the unstable manifold in the window region does not destroy the structure of the stable manifold and the trajectory was absorbed into the sink. Due to homoclinicity, the orbit is expelled from the sink to an unstable manifold, but the perturbation again put the orbit to the sink and the periodicity is realized. In the control of spiral chaos, tangency of the homoclinic orbits predicted by the Newhouse's theorem was applied. In the control of the double scroll, heteroclinic tangency accompanied by the shift in the $z$ direction was utilized.  

This kind of control could be applied to the communication systems by defining an appropriate coding function to the binary symbol sequence corresponding to rotations around $P_+$ or $P_-$\cite{hgo93}.  We observed that the number of cycles around the two fixed points depends on the detailed adjustment of the capacitance and inductance. The shape of the pulse which is not a exact delta function modifies the destination of the orbit in the region near the fixed point. Various possible controlled double scroll due to heteroclinic tangency of Shil'nikov chaos and existence of several windows region in the 1-d bifurcation diagram allows encoding characters on each cyclic oscillation and apply to communication system. 

The chaos control method using the information of the 1-d bifurcation diagram\cite{niiya06} can be applied to other systems like R\"ossler equation\cite{roes79} 
and Langford equation.\cite{lang84}. 

\leftline{\bf Acknowledgements}
We thank Hideo Nakajima for drawing our attention to \cite{kaw84} and valuable information.

\end{document}